# Light–matter interactions with photonic quasiparticles

Nicholas Rivera[1], Ido Kaminer[2]

Interactions between light and matter play an instrumental role in many fields of science, giving rise to important applications in spectroscopy, sensing, quantum information processing, and lasers. In most of these applications, light is considered in terms of electromagnetic plane waves that propagate at the speed of light in vacuum. As a result, light–matter interactions can usually be treated as very weak, and captured at the lowest order in quantum electrodynamics (QED). However, recent progress in coupling photons to material quasiparticles (e.g., plasmons, phonons, and excitons) forces us to generalize the way we picture the photon at the core of every light–matter interaction. In this new picture, the photon, now of partly matter-character, can have greatly different polarization and dispersion, and be confined to the scale of a few nanometers. Such *photonic quasiparticles* enable a wealth of light–matter interaction phenomena that could not have been observed before, both in interactions with *bound electrons* and with *free electrons*. This Review focuses on exciting theoretical and experimental developments in realizing new light–matter interactions with photonic quasiparticles. As just a few examples, we discuss how photonic quasiparticles enable room-temperature strong coupling, ultrafast "forbidden" transitions in atoms, and new applications of the Cherenkov effect, as well as breakthroughs in ultrafast electron microscopy and new concepts for compact X-ray sources.

*Key points:*

- *Photonic quasiparticles are quantized time-harmonic solutions of Maxwell's equations in an arbitrary inhomogeneous, dispersive, and possibly nonlocal medium. Surface plasmon polaritons, phonon polaritons, exciton polaritons, and all other polaritons are examples of photonic quasiparticles. Moreover, photons in cavities, localized and bulk plasmons, and even acoustic phonons are also special cases of photonic quasiparticles.*

- *Certain photonic quasiparticles can confine electromagnetic fields down to dimensions much smaller than the wavelength of a photon. Specifically, polaritons in 2D materials such as graphene and hexagonal boron nitride allow simultaneously high confinement and low optical losses.*

- *Macroscopic quantum electrodynamics (MQED) prescribes the quantization of the photonic quasiparticles in an arbitrary medium, and can describe the interaction of any photonic quasiparticle with any type of quantum matter (i.e., arbitrary emitter) in terms of its elementary emission and absorption processes.*

- *For bound-electron emitters, the confinement of photonic quasiparticles enables ultrafast spontaneous emission and few-molecule strong coupling, as well as possibilities of enabling new phenomena such as forbidden transitions and multiphoton spontaneous emission.*

- *For free-electron emitters, photonic quasiparticles enable novel applications of the Cherenkov effect in particle detectors, as well as new concepts for compact X-ray sources and new applications in ultrafast electron microscopy.*

1. Department of Physics, Massachusetts Institute of Technology, Cambridge, MA, United States. 2. Department of Electrical Engineering and Solid State Institute, Technion – Israel Institute of Technology, Haifa, Israel. e-mail: nrivera@mit.edu, kaminer@technion.ac.il

Interactions between light and matter play a crucial role in science and technology. The emission and absorption of light – by bound electrons in atoms, molecules, and solids, as well as by free electrons – form the direct basis for technologies both mature and nascent. Examples include modern spectroscopy, lasers, X-ray sources, LEDs, photo-diodes, solar cells, high-energy particle detectors, and advanced microscopy methods. Light–matter interactions are fundamentally quantum electrodynamical, and in many cases, are described as quantum transitions by electrons, accompanied by the emission, absorption, or scattering of quanta of the electromagnetic field in vacuum (photons). The theory describing photons and their interaction with electrons is nearly as old as quantum mechanics itself, and was first formulated by Dirac in 1927 [Dirac1927], with an elegant re-formulation (still used today) by Fermi in 1932 [Fermi1932]. Traditionally, it has been sufficient to describe the electromagnetic quanta as (a) composed of plane waves traveling at the speed of light and (b) having a wavelength much longer than the typical size scales of electron wavefunctions in atoms, molecules, and solids.

This traditional understanding is challenged by recent experiments using near-field microscopes to couple to polaritons in van der Waals materials, as well as recent experiments confining light in nano-gaps between metals. In particular, it is now feasible to couple light to extremely confined electromagnetic fields. Such fields – which can be plasmonic, phononic, excitonic, or even magnonic in nature – can be manipulated in many of the same ways as photons. Their close similarity to photons motivates their consideration as part of a more general concept, called *photonic quasiparticles* (Fig. 1). A photonic quasiparticle, which fundamentally arises as a quantized solution to Maxwell's equations in a medium, is a broad concept that includes not only polaritons, but also photons in vacuum and homogeneous media, photons in cavities and photonic crystals, and even, excitations that seem fundamentally non-photonic, such as bulk plasmons and bulk phonons. As such, these quasiparticles generally differ from photons in vacuum in several key respects like polarization, confinement, and dispersion. When considering how these excitations are absorbed and emitted by electrons (what we call "light–matter interactions"), one finds that these differences enable many phenomena that are difficult or even impossible to realize with photons in free space.

In systems of bound electrons (e.g., in atoms, molecules, or solids), the confinement of photonic quasiparticles strongly enhances the intrinsic coupling between these electrons and the quantized electromagnetic field. This is because the energy of the quasiparticle, $\hbar\omega$, is confined over a very small volume, leading to correspondingly strong quantized electric and magnetic fields. The enhanced coupling gives rise to greatly enhanced spontaneous emission by excited electrons. For sufficiently confined photonic quasiparticles, the enhanced coupling is strong enough to enable coherent and reversible energy exchange between the electron and the electromagnetic field. The other important effect arising from confinement is the possibility of breaking conventional selection rules governing the types of electronic transitions that can occur. In sum, these effects may enable brighter single-photon sources, highly sensitive sensing and spectroscopy platforms, and potentially even new sources of entangled quasiparticles.

Meanwhile, in systems of free electrons, the spectral and directional properties of spontaneously emitted photonic quasiparticles are sensitive to the dispersion relation of the photonic quasiparticle. Controlling the dispersion relations by using structured media – as photonic crystals, optical nanostructures, or highly confined polaritons – allows one to control "at will" the properties of light emission based on the electron energy. Importantly, the delocalized quantum wave nature of free electrons gives additional opportunities to control light–matter interactions by shaping electron wavefunctions. For example, one can shape the wavefunction to display symmetries which are compatible (or incompatible) with the symmetry of the photonic quasiparticle field, thus leveraging selection rules to control the possible interactions. Additional important effects appear when electrons interact with *strong fields* of photonic quasiparticles, which enable coherent energy exchange by means of absorption and stimulated emission. In sum, these effects may enable new and enhanced particle detection schemes, compact light sources from infrared to even X-ray frequencies, and breakthrough platforms for electron microscopy with nanometer and femtosecond resolution.

Although free and bound electron phenomena at first appear unrelated, and are typically connected to different fields of research, it is possible, and even illuminating, to take a unified view of these phenomena.

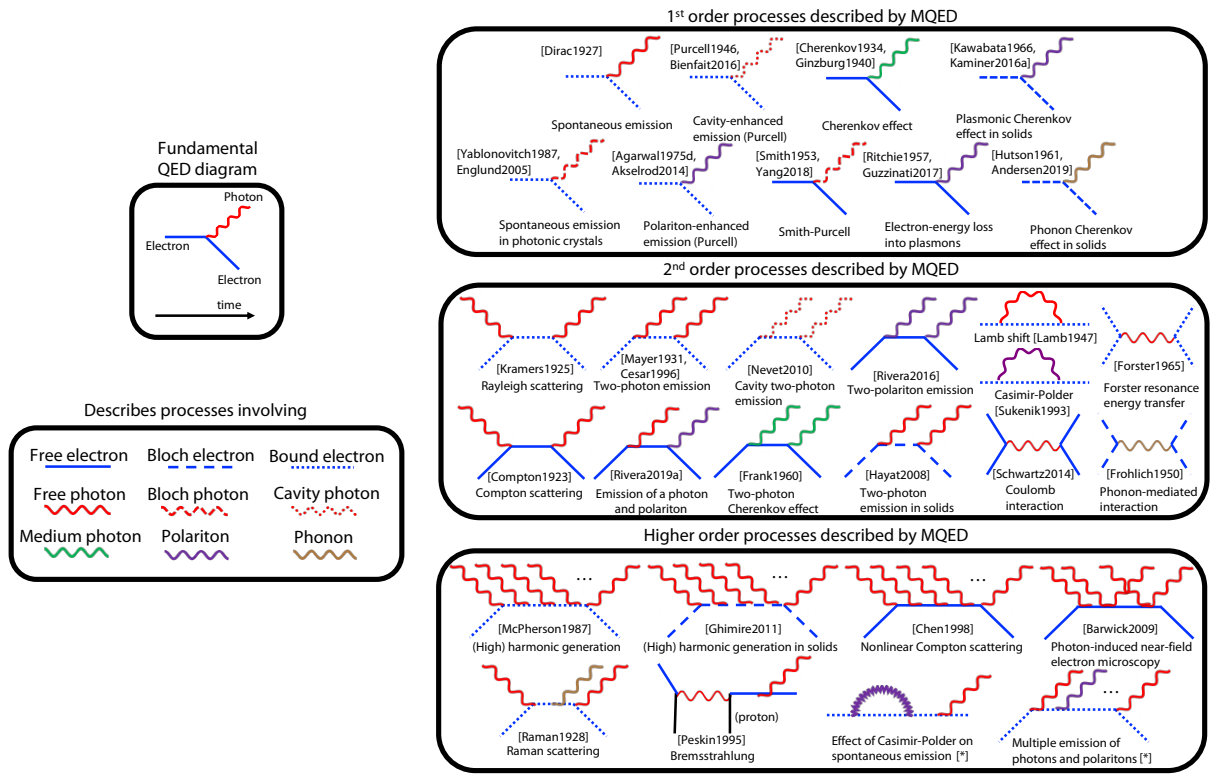

**Figure 1: Diagrammatic representation of physical processes contained within macroscopic QED (MQED)**, as they pertain to different types of matter (bound, free, and Bloch electrons), as well as different types of photonic quasiparticles (photons, photons in a homogeneous medium, photonic crystal photons, polaritons (plasmon, phonon, exciton, magnon), and even pure phonons. Processes with no standard or recent reference associated them are marked with a [*]. Each MQED diagram corresponds to a different, sometimes known phenomenon, while others correspond to phenomena which have thus far not been explored. Note that while we represent mostly spontaneous emission effects here, all spontaneous processes also have stimulated processes, as well as absorption (inverse) processes associated with them. For example, corresponding to the Cherenkov effect is the inverse Cherenkov effect, where an emitter absorbs a photon in a medium instead of emitting it. We also note here that in some cases, the emitted quasiparticle has a vacuum far-field component, leading to other effects. For example, a plasmon emitted by an electron can couple to the far-field in nanoparticles, as a mechanism of cathodoluminescence. Or a medium photon associated with an interface can have a vacuum component, leading to transition radiation.

The crux of this unified view is a systematic classification of the types of interactions that can happen between arbitrary electronic systems and arbitrary photonic quasiparticles. This classification is shown in Fig. 1, where we represent different types of elementary light–matter interaction processes between electrons and photonic quasiparticles in terms of Feynman diagrams. These diagrammatic representations emerge naturally from *macroscopic quantum electrodynamics* (MQED), which describes the interaction of electrons with electromagnetic fields in materials. An especially useful contribution from MQED that we will present in this Review is the quantization of the electromagnetic fields associated with photonic quasiparticles in terms of (classical) solutions of the macroscopic Maxwell equations in a medium.

As can be seen, changing the type of electron or the type of photonic quasiparticle in a particular Feynman diagram leads to fundamentally different phenomena, often seen as disparate physical effects. For example, spontaneous emission by atoms and molecules is loosely analogous to the Cherenkov radiation by free electrons, both being single-photonic-quasiparticle spontaneous emission processes; the Cherenkov effect is analogous to phonon amplification phenomena by electrons in solids solids, being governed by similar energy-momentum conservation rules; the phenomenon of photon-induced near-field electron microscopy is analogous to Rabi oscillations in cavity QED; high harmonic generation by bound electrons is analogous to nonlinear Compton scattering in free electrons. This line of thinking enables knowledge-transfer between different light–matter effects. Ultimately, this perspective enables one to predict and study new types of interactions that have yet to be explored.

Our Review aims to provide details to the picture painted above, by elaborating on the exciting recent theoretical and experimental developments in the field of light–matter interactions in nanophotonics, unifying the different phenomena when possible. The field of light–matter interactions in nanophotonics is broad in scope, and involves many important topics that we touch here only briefly, for which the following representative reviews provide further insight: polaritons in van der Waals materials [Basov2016, Low2017, Caldwell2019], plasmonic nano-gaps [Baumberg2019], quantum plasmonics [Tame2013], enhanced spontaneous emission phenomena [Pelton2015], strong coupling physics [Torma2014, Forn-Diaz2019, Kockum2019], electron-beam spectroscopy [Polman2019], and the theory of macroscopic quantum electrodynamics [Scheel2008].

### 1. Electromagnetic excitations as "photonic quasiparticles"

*1.1 Types of photonic quasiparticles*

*A photonic quasiparticle* is a quantized excitation of an electromagnetic mode also called "a photon of a medium" [Ginzburg1987, Lifshitz2013]. The mode is formally a time-harmonic solution to Maxwell's equations with frequency $\omega$ in an arbitrary medium, subject to boundary conditions. The electromagnetic mode corresponding to this quantized excitation is normalized such that the electromagnetic energy in a single-quasiparticle state is $\hbar\omega$, and its polarization and field-distribution are governed solely by the response functions of the medium: the dielectric permittivity $\epsilon$ and magnetic permeability $\mu$. In Fig. 2 (top), we show some of the types of microscopic phenomena that can contribute to the response functions, such as free-electrons (in metals), bound electrons (in simple insulators like glass), optical phonons (in polar dielectrics), magnons (in ferro- and anti-ferromagnets), and excitons (in semiconductors). These *microscopic* phenomena define the frequency-dependence of the *macroscopic* response functions of the material. Different materials, as well as different geometries of the materials, lead to qualitatively different kinds of photonic quasiparticles, as shown in Fig. 2 (bottom). Let us now consider a systematic classification of the different types of photonic quasiparticles that exist, based on dimensionality, with an eye towards the effects in light–matter interactions enabled by each type of quasiparticle.

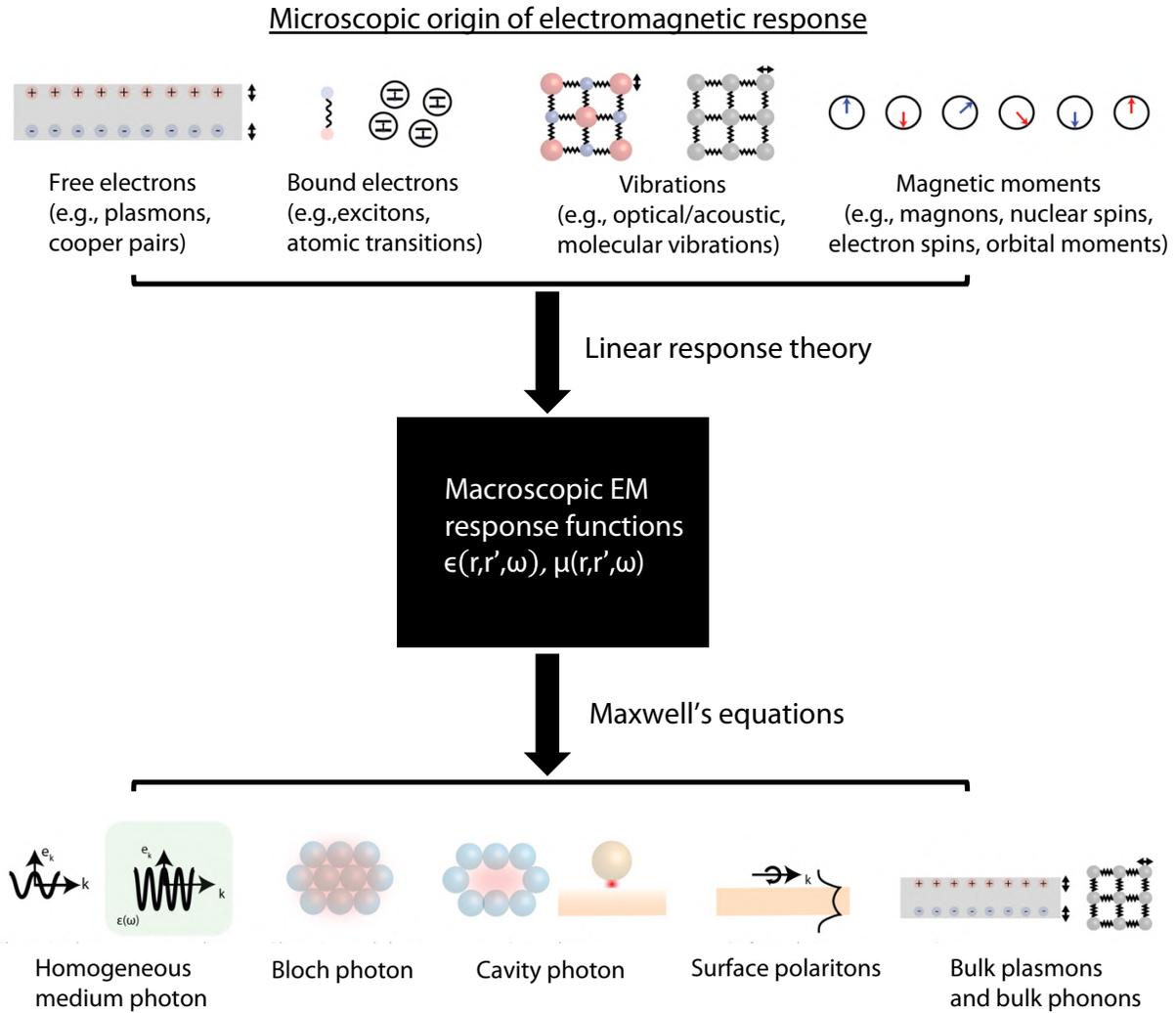

**Figure 2: Photonic quasiparticles.** The electromagnetic interactions of bound and free electrons with materials can be unified into a single framework. In this framework, the microscopic origin of the electromagnetic excitations (top) "collapses" into a spatially and temporally dispersive dielectric permittivity and magnetic permeability (middle), which is essentially a black box. The linear electromagnetic response functions can be calculated from the microscopic properties through linear response theory. The material properties, combined with material geometry, give rise to different types of photonic quasiparticles (bottom). Examples of these limits include photons in vacuum and homogeneous media, photonic crystal photons, cavity photons, surface polaritons, and, even bulk plasmon and phonon excitations.

*3D translationally-invariant photonic quasiparticles.* The simplest examples of photonic quasiparticles are those in a 3D translation-invariant bulk, which supports propagating plane waves that are characterized by their frequency, momentum, propagation lifetime, and polarization. The polarization is transverse to the electric displacement **D** or magnetic field **H**, unless $\epsilon(k,\omega) = 0$ or $\mu(k,\omega) = 0$ respectively. If $\epsilon(k,\omega) = 0$ or $\mu(k,\omega) = 0$, longitudinal modes of Maxwell's equations are allowed, like bulk plasmons and phonons, or bulk magnons in the magnetic case. Even in a homogeneous medium, there exist several distinct kinds of photonic quasiparticles, which include photons in vacuum, photons in a transparent medium (e.g., glass), bulk polaritons, and their quasi-static analogues (e.g., bulk plasmons, bulk phonons, etc.). A key difference between these photonic quasiparticles and photons in vacuum is that some have phase velocities below the speed of light $c$, with bulk plasmons and phonons having velocities far below the speed of light. These reduced phase velocities enable phenomena such as radiation from uniformly moving charges, e.g., the Cherenkov effect in a dielectric medium [Friedman1988], bulk plasmon emission processes measured in electron-energy loss spectroscopy [deAbajo2010], and even phonon emission processes by electrons in solids [Giustino2017].

*2D & 1D translationally-invariant photonic quasiparticles.* 2D translation-invariant systems include thin films, slabs, interfaces between two semi-infinite materials, multilayer stacks, and 2D materials. Such systems support several kinds of photonic quasiparticles, including waveguide modes in dielectric slab waveguides and hyperbolic media (such as hexagonal boron nitride), and confined surface modes that evanescently decay from the surface (e.g., surface plasmon polaritons and surface phonon polaritons in conventional media). Because the class of 2D translationally invariant photonic quasiparticles includes both thick and thin films, some examples of photonic quasiparticles such as slab waveguide modes and hyperbolic surface phonon polaritons could be considered as being both surface (due to their evanescent tails) and bulk (due to their propagation in the medium). 2D translationally invariant modes are characterized by their frequency, in-plane momentum, propagation lifetime, and polarization. Waveguide modes can further have a discrete mode order that determines their out-of-plane field distribution.

From the standpoint of light–matter interactions, the polarization and dispersion of 2D translationally invariant systems lead to many effects that do not occur with photons in vacuum. For example, evanescent modes can have circular polarization in the plane perpendicular to their magnetic field. The chirality is locked to the direction of propagation (spin-momentum locking [Bliokh2015]), so that right-moving and left-moving waves have opposite chirality. Thus, an emitter with a circularly polarized transition dipole moment can only emit waves in one direction, as waves in the opposite direction have zero overlap with the dipole [Lodahl2017]. In another example related to polarization, because the polarization of a surface mode is partially out-of-plane, a surface mode overlaps well with a vertically oriented transition dipole associated with a planar emitter such as excitons in a transition metal dichalcogenide [Zhou2017]. This lies in contrast to free-space, where the transversality of the electromagnetic wave implies that vertically oriented dipoles cannot emit at normal incidence (zero overlap), rendering them optically dark and difficult to detect in the far-field. This enables one to perform spectroscopy with dark excitons based on surface plasmons [Zhou2017].

Another key difference in light–matter interactions comes from the fact that systems with negative permittivity (polaritonic systems), support surface modes with wavelengths far smaller than that of a photon of the same frequency [Jablan2009, Fei2012, Chen2012, Dai2014, Caldwell2014, Lundeberg2017, Iranzo2018, Ni2018], corresponding to a highly confined out-of-plane field. Such confinement leads to a very high local density of electromagnetic states, and consequently, quantum emitters in the vicinity of these modes can interact quite strongly with them, manifesting in enhanced spontaneous emission, as well as breakdown of selection rules associated with the dipole approximation. These effects are elaborated in Section 2. Experimentally, such quasiparticles have been leveraged for high-resolution nano-imaging of electrons in solids [Basov2016, McLeod2017, McLeod2019], sensitive sensors of vibrational transitions in molecules [Rodridgo2015, Autore2018], and enhanced interactions with quantum emitters [Tielrooij2015]. Similar conclusions to those discussed for 2D modes also apply in 1D translation-invariant systems (e.g., fibers and other waveguides) [LeKien2004, Yan2013, Junge2013].

Importantly, the photonic quasiparticle concept also applies in systems with discrete translation invariance (periodic systems), in any dimension, where it includes photonic crystal modes (Bloch photons) [Joannopoulos2008].

*0D translationally-invariant photonic quasiparticles.* Systems with 0D translation-invariance (i.e., that fully break translation-invariance) support localized cavity modes, a distinct type of photonic quasiparticle characterized by its frequency, lifetime, polarization, and field distribution (setting its *mode volume*). In particular, cavities with high quality factors support photonic quasiparticles such as whispering-gallery modes [Spillane2002, Armani2003, Vahala2003] and photonic crystal defect modes [Akahane2003, Song2005], used for example for enhanced sensors and for low-threshold laser interactions. Of importance for this review are cavities with ultra-high quality factors [Spillane2002, Armani2003, Vahala2003, Song2005] and ultra-small mode volumes (e.g., plasmonic and phonon-polaritonic cavities). Both can enable enhanced spontaneous emission due to the concomitant enhancement of the local-density of states [Tanaka2010, Akselrod2014, Chikkaraddy2016, Benz2016]. This feature is similar to highly confined propagating quasiparticles in 1D and 2D. One major difference in 0D systems is that boundary conditions force a quasi-discrete spectrum for the modes, leading to sharp spectral peaks in the local density of states – in contrast to systems with propagating modes, whose spectrum is continuous. Qualitatively, the interaction of quantum emitters with a discrete mode is quite different from that with continuum modes. In the former case, the system resembles two coupled oscillators, allowing new normal modes of the emitter and cavity mode to form (strong coupling). In the latter case, a discrete emitter undergoes irreversible decay into the continuum (enhanced spontaneous emission), provided that the coupling is not too strong.

Special types of photonic quasiparticles that do not fit as neatly into the above categorization can be constructed by superposition of extended modes, which breaks their translation invariance and effectively localizes them. For example, a cylindrically symmetric superposition of surface plasmons creates plasmon vortices characterized by an integral orbital angular momentum (OAM) quantum number. Such 2D vortices have been observed on various metal-insulator surfaces [David2015, Spektor2017, Spektor2019] and predicted in graphene and hexagonal boron nitride [Du2014]. More advanced superpositions can be used to create arrays of vortices with topological features [Tsseses2018]. From the standpoint of light–matter interactions, photonic quasiparticles with OAM are interesting because when an electron absorbs or emits such a quasiparticle, its angular momentum must change by the OAM of the quasiparticle (provided the emitter and vortex are concentric) [Babiker2002, Schmiegelow2016, Machado2018]. Controlling dynamics with OAM-possessing photonic quasiparticles also applies in the case of free-electron absorption and stimulated emission [Cai2018, Vanacore2019].

*Example of photonic quasiparticles: Polaritons in van der Waals materials*. An important example of photonic quasiparticles are polaritons in van der Waals and 2D materials – primarily plasmon and phonon polaritons. They are of great recent interest because unlike photons in conventional dielectrics, they can be confined to volumes over a million times smaller than that of a diffraction-limited photon in vacuum, which can enable many new effects in light–matter interactions, as well as enhanced sensors, and enhanced optical nonlinearities. The basic physics of polaritons is well-described in recent reviews (e.g., [Caldwell2015, Basov2016, Low2017]); our focus is on their unique light–matter interactions, emphasizing the key similarities and differences to free-space photons.

In Fig. 3, we summarize recent experiments probing polaritons in thin films and 2D materials, demonstrating that optics can be performed with polaritons. In planar slabs of polaritonic materials, the polariton has an in-plane wavevector much larger than the wavevector of a photon at the same frequency. Due to the continuous translational symmetry of the slab, photons incident from the far-field cannot couple directly to the slab (ignoring the edges of the slab), necessitating the use of a coupling element that provides momentum to the incident photon, enabling momentum conservation. The most common examples are sharp tips and gratings. A sharp tip fully breaks in-plane translation symmetry, allowing an optical far-field to launch polaritonic waves from the tip, as is central to methods like scanning near-field optical microscopy (SNOM). Such methods are used extensively to measure the complex dispersion relation (wavenumber and propagation length), as well as the spatial distribution of the electric field, in various polaritonic systems: plasmons in graphene [Fei2012, Chen2012, Woessner2015, Lundeberg2017, Ni2018], phonon polaritons in hexagonal boron nitride (thin films and monolayers) [Dai2014, Li2015, Yoxall2015, Giles2018], exciton polaritons in molybdenum selenide [Hu2017, Mrejen2019], and newer materials such as hyperbolic phonon polaritons in molybdenum trioxide [Li2016, Ma2018].

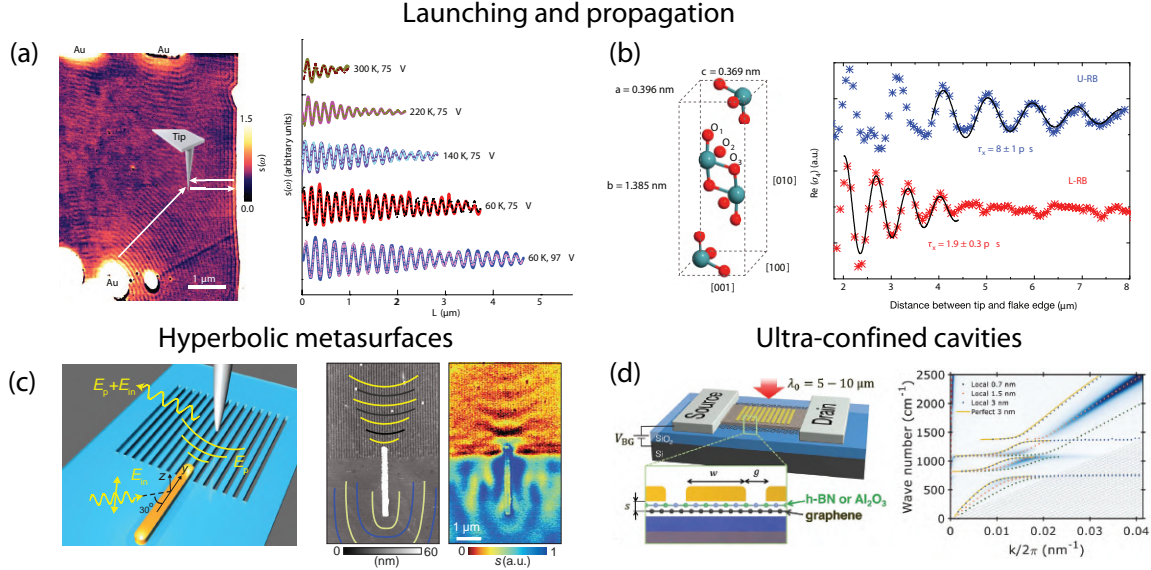

**Figure 3: Optics with photonic quasiparticles.** Photonic manipulations (optical excitation, propagation, coupling to structures, and detection of electromagnetic fields), with highly confined polaritons. (a) Propagation of plasmons in graphene at low temperatures, such that the losses are very low [Ni2018]. (b) Propagation of phonon polaritons in newly discovered, in-plane hyperbolic material $MoO_3$ [Ma2018]. (c) Launching hyperbolic phonon polariton waves by an antenna structure, in some sense performing a similar role to the metallic tip in scanning probe microscopy, but allowing strong control over the phase fronts of the polaritonic radiation [Li2018]. (d) Plasmons in a doped graphene situated a nanometer away from a gold grating structure, allowing for confinement of the electromagnetic field on the scale of a few atoms [Iranzo2018].

Figs. 3a and 3b show direct examples of the highly confined nature of the polaritons. In Fig. 3a, showing a recent example with plasmons in graphene, the plasmon is measured to have a wavelength over 100 times smaller than the wavelength of a photon in vacuum. This is a key difference from photonic quasiparticles in all-dielectric systems. Fig. 3a shows the exceptionally long lifetime that can be achieved with graphene plasmons (roughly 130 optical cycles), which was facilitated by operating at low temperature to suppress losses related to acoustic phonon coupling. The combination of high confinement and low loss is instrumental not only in envisioning optical components based on the propagation of plasmons, but more generally in enhancing light–matter interactions with quantum emitters. Such enhancement depend on the local density of optical states that increase with high confinement and low loss. In Fig. 3b, we show a recent SNOM of highly-confined phonon-polaritons in molybdenum trioxide, whose wave-fronts demonstrate the hyperbolic nature of the polaritons in this material [Ma2018], potentially enabling new platforms for hyperbolic optics in the mid-infrared spectral region.

Various antenna structures can also be used to assist the coupling of light into the photonic quasiparticle mode, as in Fig. 3d, where a gold rod is used to launch phonon polaritons in gratings of hexagonal boron nitride, which act as a hyperbolic metasurface. Due to the opposite signs of in- and out-of-plane permittivities, wavefronts launched from the rod exhibit spatial propagation profile in a clear signature of hyperbolicity [Li2018], enabling one to study light–matter interactions between emitters and hyperbolic quasiparticles. Similar methods using antennas have also been used to launch graphene plasmons [Alonso2014].

The interaction with polaritons can also be facilitated with grating structures, as in Fig. 3c, where a grating-cavity consisting of a gold-grating atop a gold mirror sandwiches boron-nitride-encapsulated graphene [Iranzo2018] (encapsulation improves the lifetime of the plasmon, due to suppression of coupling to phonons [Principi2014]). The grating couples far field light into the cavity-enhanced graphene plasmons that benefits from the very high reflectivity of gold at the mid-infrared wavelength. While the lifetime is modest, being on the order of ten optical cycles, what is remarkable here is that this cavity achieves out-of-plane confinement of the graphene plasmon to the scale of 1 nm, representing the smallest mode-volume graphene plasmon ever measured, with an estimated mode volume on the order of $10^{-9}\lambda_0^3$. Such small volumes could enable extremely non-perturbative interactions between light and matter.

*Example of photonic quasiparticles: Plasmons in metallic nanogaps.* Another important class of photonic quasiparticles in this Review are plasmons in the "conventional" noble metals such as gold and silver. Confined surface plasmons can be supported in these systems based on thin films and metal-insulator-metal structures. We focus particularly on localized plasmonic cavities as they have been the workhorse of recent experiments in strong quantum light–matter interactions. We leave detailed discussion of the electromagnetic physics of these cavities to dedicated reviews as [Baumberg2019]. Plasmonic nano-gaps typically involve the geometry of a metallic nano-particle (such as a nano-sphere or nano-disk) separated from a planar metal film by a very small gap, which can be on the order of 1 nm. This geometry is referred to as a nanoparticle-on-mirror (sometimes abbreviated as NPoM) geometry or as a plasmonic nanogap cavity.

Recent experiments have demonstrated the existence of these strongly confined cavity modes based on nanogaps as "large" as 5 nm [Akselrod2014], moving recently to sub-nanometer sizes [Chikkaraddy2016]. A striking recent example of this geometry at its ultimate limit is that of the picocavity [Benz2016], which leads to strong field enhancements in a single atom protrusion from a nanoparticle, explained in terms of a type of lightning-rod effect. It is instrumental to note the values of the polarization, lifetime, and mode volume of these types of modes: the polarization is primarily perpendicular to the interfaces, the lifetimes tend to be roughly one to ten optical cycles (with potential improvements coming from hybrid dielectric-metal geometries [Yang2017]), and the mode volumes have been estimated to be below 1 nm$^3$. The extreme confinement of such cavities makes effects related to spatial nonlocality particularly strong [Yang2019]; such effects are of considerable importance as they are likely to provide fundamental limitations on applications of nanophotonics and light–matter interactions.

### *1.2 Quantum electrodynamics with photonic quasiparticles*

Although the examples above have thus far been understandable from solutions of the classical Maxwell equations, experiments have also demonstrated the underlying quantum nature of the electromagnetic fields of these photonic quasiparticles through quantum optical measurements. Many of these experiments have been in the context of plasmonics. For example, quantum statistics of plasmons were demonstrated [Akimov2007], along with plasmonic preservation of photon entanglement [Altewischer2002] and two-plasmon quantum interference in a Hong-Ou-Mandel experiment [Fakonas2014].

Perhaps more simply, phenomena like spontaneous emission in any material system already call for a quantized description of the electromagnetic fields associated with each type of photonic quasiparticle. The key theoretical framework that prescribes the quantization of *any* photonic quasiparticle and the interactions of these quasiparticles with emitters is called *macroscopic quantum electrodynamics (MQED)* [Knoll2000, Scheel2008, Philbin2010]. It is "macroscopic" because it treats the photonic quasiparticles as being governed by the macroscopic Maxwell equations. i.e., MQED treats the medium in terms of permittivities and permeabilities, taking the microscopic charges and currents in the medium as continuous. As an important point of terminology, since MQED handles the quantization of the EM field in any linear medium, its special cases cover all the effects of "other QEDs" in the literature such as cavity, circuit, waveguide, photonic-crystal, and plasmonic QED.

*The quantization of photonic quasiparticles*. Pedagogically, it is useful to explain the quantization of the electromagnetic field in two steps: in the first, the fields are quantized in ideal, lossless materials, and in the second, they are quantized in arbitrary absorbing materials. Quantization of electromagnetic fields in lossless materials is long-known, as exposited in [Knoll1987, Glauber1991]. For most cases of interest, the lossless case describes very well the essential physics of the emission and absorption of photonic quasiparticles by emitters, bound or free. With this in mind, we first describe the quantization in lossless materials in a constructive way that introduces the terminology to be used more generally later. The absorbing case is presented in Box 1.

In lossless and non-dispersive materials, we may represent an electromagnetic field operator (such as the vector potential $\boldsymbol{A}(\boldsymbol{r}, t)$ in the Heisenberg picture) in terms of an expansion over time-harmonic modes $\boldsymbol{F}_n(\boldsymbol{r})e^{-i\omega_n t}$. These modes capture all of the details of the frequency, polarization, and field distributions of the photonic quasiparticles described in the previous section (e.g., dispersion relations, polarization properties, and field distributions). In this expansion, each mode, $n$, is associated with a quantum harmonic oscillator [Glauber1991], with associated creation ($a_n^\dagger$) and annihilation ($a_n$) operators, satisfying the canonical bosonic commutation relations: $[a_m, a_n] = [a_m^\dagger, a_n^\dagger] = 0$ and $[a_m, a_n^\dagger] = \delta_{mn}$. The resulting vector potential takes the form:

$$\boldsymbol{A}(\boldsymbol{r}) = \sum_n \sqrt{\frac{\hbar}{2\epsilon_0 \omega_n}} (\boldsymbol{F}_n(\boldsymbol{r}) a_n + \boldsymbol{F}_n^*(\boldsymbol{r}) a_n^\dagger). \qquad (1)$$

For a non-magnetic medium, the mode satisfies $\nabla \times \nabla \times \boldsymbol{F}_n(\boldsymbol{r}) = \epsilon(\boldsymbol{r}) k_n^2 \boldsymbol{F}_n(\boldsymbol{r})$, with $k_n = \omega_n/c$ [Joannopoulos2001] and are normalized such that $\int d\boldsymbol{r} \ \epsilon(\boldsymbol{r}) |\boldsymbol{F}_n(\boldsymbol{r})|^2 = 1$. This normalization makes it so that a one-photon state has an electromagnetic energy of $\hbar\omega$ relative to the vacuum state. This mode expansion is immediately applicable to QED phenomena in low-loss cavities, waveguides, and photonic crystals. In practice, mode expansions can also be used in the case of dispersive materials such as the polaritonic materials presented in Fig. 3 (provided the modes kept in the mode expansion are of low loss), by changing the normalization condition (as in Table 1 of Box 1). The adjusted normalization condition arises because the energy of the quanta in a dispersive system is governed by the Brillouin energy density formula for dispersive materials [Landau2013ECM, Archambault2010]. We mention here that these mode expansions are not valid at all frequencies in dispersive materials, because regions of high loss generally exist, particularly in polaritonic materials. The examples shown in Fig. 3 are chosen intentionally to coincide with low enough loss.

*Quantum interactions between emitters and photonic quasiparticles.* Once photonic quasiparticles are quantized, we now quantitatively describe how these quasiparticles interact with bound and free electrons (collectively referred to as emitters). For this purpose, we consider transitions between electronic states of the emitter that are accompanied by the emission, absorption, or scattering of single or multiple photonic quasiparticles (either real, as in spontaneous emission, or virtual, as in Lamb shifts/Casimir-Polder forces). Examples of these processes for bound and free electrons were shown in Fig. 1, with examples of the relevant photonic quasiparticles shown in Fig. 2 and Section 1.1.

In non-relativistic bound electron systems, these transitions are described by the *Pauli-Schrodinger Hamiltonian*, or a suitably approximated version of it (see Box 2). In free-electron systems (relativistic or non-relativistic), the transitions are governed by the Dirac Hamiltonian, in cases where electron spin is important, or the Klein-Gordon Hamiltonian, where it is not (see Box 3). In both cases, the transitions are described by a term proportional to $\mathbf{A} \cdot \mathbf{v}$, provided that the electron does not change its energy significantly upon emission or absorption. This term couples the quantized vector potential to the velocity of the electron, described in terms of its momentum by $\mathbf{v} = (\mathbf{p} - q\mathbf{A})/m$, with $q$ the electric charge and $m$ the mass of the electron.

The key element in any calculation of light–matter processes with photonic quasiparticles is the rate of transition between some initial quantum state $i$ and some final quantum state $f$. See Fig. 1 for examples of initial and final states corresponding to known light–matter interaction processes. This rate of transitions at arbitrary order in the perturbation can be found by an iterative procedure [Landau2013QM]. The most commonly occurring cases are the transition rates at first (1) and second (2) order in QED, which are respectively given as $\Gamma^{(1)} = (2\pi/\hbar)|V_{fi}|^2 \delta(E_f - E_i)$, and $\Gamma^{(2)} = (2\pi/\hbar) \lim_{\eta \to 0} \left|\sum_n \frac{V_{fn}V_{ni}}{E_i - E_n + i\eta}\right|^2 \delta(E_f - E_i)$. Here, $V_{ab} \equiv \langle a|V|b \rangle$, with $V = -q\mathbf{A} \cdot \mathbf{v}$ being the interaction Hamiltonian of QED, and $n$ denotes an intermediate (virtual) state to be summed over. The delta functions express the conservation of energy between initial and final states. Energy shifts associated with emission and re-absorption of virtual photonic quasiparticles (Lamb shifts, Casimir-Polder forces) can be described by time-independent perturbation theory, with the shift in energy $\delta E_i$ of quantum state $i$ given as $\delta E_i = \lim_{\eta \to 0} \sum_n \frac{|V_{ni}|^2}{E_i - E_n + i\eta}$.

So far, the principles of MQED in its lossless and its lossy varieties have been used with the interaction terms above to describe a plethora of phenomena: atomic spontaneous emission of one and two photons (see e.g., [Agarwal1975, Glauber1991, Scheel2008, Rivera2016, Rivera2017]), emission from extended emitters in solids like quantum wells (e.g., [Kurman2018]), strong-coupling effects in bound emitters (e.g., [Scheel2008, Kurman2020]), cavity/circuit/waveguide/plasmonic/photonic crystal QED phenomena (e.g., [Kleppner1981, Yablonovitch1987, John1990, John1994, Zhu2000, Yao2009, Gonzalez-Tudela2013, Gonzalez-Tudela2014]), energy shifts due to virtual photon emission and absorption as the Lamb shifts/Casimir-Polder forces (e.g., [Thirunamachandran1981, Ribeiro2015]), Casimir forces (e.g., [Buhmann2007, Scheel2008, Buhmann2008]), and even phenomena associated with emission of photonic quasiparticles by ultra-relativistic electrons [Kaminer2016b, Rivera2019a], as well as electrons driven by strong external fields [Rivera2019c]. Generally, it can be used to describe any of the processes illustrated in Fig. 1.

## 2. Light–matter interactions with photonic quasiparticles: bound electron systems

The bulk of the Review discusses how the photonic quasiparticles described above are used to enhance and control the classical and quantum interactions of electromagnetic fields with electrons in atoms, molecules, solids, and even with free electrons (collectively referred to as "emitters"). For each type of emitter, it is useful to further divide the interactions by whether they are "weak-coupling" effects, such as emission, absorption, and scattering, where the perturbative description of light–matter coupling is valid, or "strong-coupling" effects, where the perturbative description is not valid. We survey both regimes below. In all cases, we consider the effects of different types of photonic quasiparticles.

### *2.1 Controlling bound electron spontaneous emission with photonic quasiparticles.*

*Spontaneous emission with photonic quasiparticles.* A key effect arising from photonic quasiparticles is that the spontaneous emission of excited emitters (bound or free) can take place by emission of a photonic quasiparticle different from a photon in vacuum. This effect, first investigated theoretically in the context of nuclear magnetic dipole emission by Edward Purcell in 1946, is today referred to as the *Purcell effect*. Quantum mechanically, spontaneous emission corresponds to a transition between an excited electron (energy $\hbar\omega_i$) with no photonic quasiparticles $|i, 0\rangle$, to a set of final emitter states (energy $\hbar\omega_f$) with one photonic quasiparticle at some mode $\{|f, 1\rangle\}$. For a fixed final electron state $f$, the emission rate $\Gamma_{fi}$ can be derived by applying Fermi's Golden Rule at first-order in time-dependent perturbation theory, using the quantized electromagnetic field of an arbitrary medium according to MQED [Rivera2016]:

$$\Gamma_{fi} = \frac{2\mu_0}{\hbar} \int d^3r \, d^3r' \, \boldsymbol{j}_{fi}^*(\boldsymbol{r}) \cdot \text{Im} \, \boldsymbol{G}(\boldsymbol{r}, \boldsymbol{r}', \omega_{fi}) \cdot \boldsymbol{j}_{fi}(\boldsymbol{r}') \approx \frac{2\mu_0 \omega_{fi}^2}{\hbar} \boldsymbol{d}_{fi}^* \cdot \text{Im} \, \boldsymbol{G}(\boldsymbol{r}, \boldsymbol{r}, \omega_{if}) \cdot \boldsymbol{d}_{fi}, \quad (2)$$

where $\omega_{if} = \omega_i - \omega_f$, and $\boldsymbol{j}_{fi}(\boldsymbol{r}) = q\psi_f^*(\boldsymbol{r})(\boldsymbol{p}/m)\psi_i(\boldsymbol{r})$, with $\psi_{i(f)}$ being the initial (final) emitter wavefunction, $q$ is the emitter charge, $m$ its mass, and $\boldsymbol{p}$ its momentum operator. $\boldsymbol{G}(\boldsymbol{r}, \boldsymbol{r}', \omega)$ is the Green's function of the Maxwell equations describing the surrounding medium. The final formula can also be expressed in terms of ratio the local density of optical states (LDOS) of the medium, $\boldsymbol{\rho}(\boldsymbol{r}, \omega_{if}) \equiv \frac{6\omega_{fi}}{\pi c^2} \text{Im} \, \boldsymbol{G}(\boldsymbol{r}, \boldsymbol{r}, \omega_{if})$, to that of the far field, $\rho_0(\omega_{if})$, via

$$\Gamma_{fi} = (\hat{\boldsymbol{d}}_{fi}^* \cdot \boldsymbol{\rho}(\boldsymbol{r}, \omega_{if}) \cdot \hat{\boldsymbol{d}}_{fi}/\rho_0(\omega_{if})) \, \Gamma_0. \quad (3)$$

where $\hat{\boldsymbol{d}}_{fi}$ is the direction of the transition dipole, and $\Gamma_0 = \frac{\omega_{if}^3 d_{fi}^2}{3\pi\epsilon_0 \hbar c^3}$ the rate of spontaneous emission into photons in vacuum [Endnote1].

The quantity $\boldsymbol{j}_{fi}$ is known as the *transition current density*, and its introduction reveals that the emission rate is, up to a factor of 2, $W_{fi}/\hbar\omega_{if}$, where $W_{fi}$ is the *classical* work done on this transition current by its own radiated field. The right-hand side of the equation holds under the *dipole approximation* (or long-wavelength approximation), i.e., that $\boldsymbol{j}_{fi}$ is localized over a scale much smaller than that of the optical field, with $\boldsymbol{d}_{fi} = \int d^3r \, q\psi_f^*(\boldsymbol{r})\boldsymbol{r}\psi_i(\boldsymbol{r})$ being the *transition dipole moment*. This formulation allows numerical simulation of the Purcell effect in complex electromagnetic geometries via classical electromagnetic simulations based on e.g., finite-element, finite-difference, or boundary-element methods. The radiated flux to each final state $f$ can be calculated by solving the classical electromagnetic problem for a dipole source $\boldsymbol{d}_{fi}$ or a more general current source $\boldsymbol{j}_{fi}(\boldsymbol{r})$, where each such source is calculated using the quantum mechanical wavefunctions. From the above equation, it can be seen that the validity of such an approach is not limited to dipole emitters but is general to any quantum emitter characterized by its transition current density.

Although the approach here makes use of MQED in lossy media, it conforms with the mode expansions of Section 1.2 by recognizing that in the lossless limit, the imaginary part of the Green's function is given by a mode expansion of the form $\text{Im}\,\boldsymbol{G}(\boldsymbol{r},\boldsymbol{r}',\omega_{if}) = \frac{\pi c^2}{2\omega_{fi}}\sum_n \boldsymbol{F}_n(\boldsymbol{r}) \otimes \boldsymbol{F}_n^*(\boldsymbol{r}')\delta(\omega_{if} - \omega_n)$ [Novotny2012], leading to a decay rate in terms of modes given by $\Gamma_{fi} = \frac{\pi}{\epsilon_0 \hbar \omega_{fi}}\sum_n |\int d^3r\,\boldsymbol{j}_{fi}^*(\boldsymbol{r})\cdot \boldsymbol{F}_n(\boldsymbol{r})|^2 \delta(\omega_{if} - \omega_n)$. In the dipole approximation, this becomes $\Gamma_{fi} \approx \frac{\pi \omega_{fi}}{\epsilon_0 \hbar}\sum_n |\boldsymbol{d}_{fi}^* \cdot \boldsymbol{F}_n(\boldsymbol{r})|^2 \delta(\omega_{if} - \omega_n)$.

*The case of a dipole emitter: the Purcell effect.* One of the most common and instructive examples of the Purcell effect involves the enhancement of spontaneous emission of a dipole emitter in an optical cavity. For a single-mode cavity, the electric field can be expressed as $\boldsymbol{E}(\boldsymbol{r},t) = (\boldsymbol{u}(\boldsymbol{r})/\sqrt{V})e^{-i\omega t - \Gamma t/2}$, with $\boldsymbol{u}(\boldsymbol{r})$ a dimensionless function dictating the spatial mode profile, $V$ the mode volume, and $\Gamma$ the decay rate of the mode. As there is an arbitrary degree of freedom in defining the mode volume versus the normalization of $\boldsymbol{u}(\boldsymbol{r})$, it can be chosen so its maximum value is 1. The imaginary part of the Green's function of this single mode can be written as a Lorentzian [Novotny2012]: $\text{Im}\,\boldsymbol{G}(\boldsymbol{r},\boldsymbol{r},\omega_{fi}) = \frac{c^2}{V}\frac{\Gamma\omega}{(\omega_{fi}^2 - \omega^2)^2 + (\Gamma\omega)^2}\boldsymbol{u}^*(\boldsymbol{r}) \otimes \boldsymbol{u}(\boldsymbol{r})$. Defining the quality factor, $Q = \omega/\Gamma$, the spontaneous emission rate on resonance ($\omega = \omega_{fi}$) immediately follows as:

$$\Gamma_{fi} = \frac{3}{4\pi^2}\frac{Q}{(V/\lambda_0^3)}\left|\widehat{\boldsymbol{d}}_{fi}\cdot \boldsymbol{u}(\boldsymbol{r})\right|^2 \Gamma_0, \tag{4}$$

with $\lambda_0 = 2\pi c/\omega_{fi}$ the photon wavelength in vacuum. Since $\Gamma_{fi}/\Gamma_0$ is proportional to the LDOS, we see immediately that the LDOS goes as $Q/V$, i.e., it is enhanced by high quality factors and small modal volumes.

When the transition dipole overlaps perfectly in polarization and is located at the maximum of the mode ($\left|\widehat{\boldsymbol{d}}_{fi}\cdot \boldsymbol{u}(\boldsymbol{r})\right|^2 = 1$), the expression coincides with Purcell's famous formula [Purcell1946]. Experiments involving the Purcell effect often have many emitters that are not located at the maximum of the mode and whose polarizations do not perfectly overlap with the field polarization – leading to less dramatic enhancements than predicted by the ideal Purcell formula. Another effect that can be appreciated from the Lorentzian dependence of the Green's function is that for an emitter far off-resonance from the cavity, $\Gamma_{fi} < \Gamma_0$, representing an inhibition of spontaneous emission [Kleppner1981].

Typically, the *Purcell factor* $F_p = \Gamma_{fi}/\Gamma_0$ is either optimized by maximizing $Q$ or by minimizing $V$. That said, spontaneous emission enhancement need not rely on a cavity, as spontaneous emission can also be enhanced for emitters coupled to waveguides or polaritonic films that support *propagating* photonic quasiparticles. In such systems, the quality factor of the propagating waves does not play the essential role it plays in cavities, because of the continuous dispersion $\omega(k)$ of the waves. However, the *confinement factor* $\eta = ck/\omega(k) = \lambda/\lambda_0$ of the modes plays the role of the mode volume, leading to strong enhancement of spontaneous emission into propagating modes that are very sub-wavelength compared to photons in vacuum. In particular, the emission into thin film modes, up to factors of order unity, scales as $\Gamma_{fi} \sim \frac{\eta^2}{(v_g/c)}\Gamma_0$, with $v_g$ the group velocity of the mode. Taking the magnitude of the group and phase velocities to be similar (to order one factors), one then has $\Gamma_{fi} \sim \eta^3 \Gamma_0$, stating that the spontaneous emission into surface modes is enhanced by the "volumetric confinement" of the polariton.

Strong Purcell enhancement can be achieved by means of a small modal volume cavity as realized in plasmonic nanogap structures [Akselrod2014] (Fig. 4a). In this experiment, the authors demonstrated directly by time-resolved fluorescence measurements how dye molecules sitting in a few-nm gap between a gold nanocube and gold film (a nanoparticle-on-mirror geometry) emit into the cavity mode far faster than they emit directly into the far-field. This particular experiment shows an increase in the spontaneous emission rate in excess of 1,000, with other experiments in the same geometry showing fluorescence enhancements of 30,000 [Rose2013]. Similar enhancements have been proposed with e.g. polaritons in van der Waals materials, such as graphene plasmons or phonon polaritons in hBN. The first to do so was by Koppens *et. al.*, predicting spontaneous emission rate enhancements of one-million-fold in doped nano-disk cavities [Koppens2011]. The $Q/V$ ratio needed for this level of enhancement has been inferred experimentally in a few graphene-plasmonic systems, and in phonon-polaritonic systems based on hexagonal boron nitride and silicon carbide [e.g., Caldwell2013, Nitikin2016, Iranzo2018].

So far, such enormous enhancements have yet to be demonstrated, perhaps due to the fact that a suitable emitter has yet to be identified that can be integrated with graphene plasmons, although some recent works along this direction are promising [Schadler2019]. To that end, experiments with erbium atoms near doped graphene surfaces showed that that relaxation rate of excited erbium atoms was strongly modified in the vicinity of graphene. That work indirectly showed enhancement factor on the order of 1,000, and dependence of the relaxation rate on the doping level in graphene, which enabled several different regimes of decay into electron-hole pairs, plasmons, and photons [Tielrooij2015].

<u>*Novel spontaneous emission processes enabled by photonic quasiparticles.*</u> Transitions associated with emission or absorption are typically associated with emission of a single photonic quasiparticle (per emitter) and typically obey dipole selection rules. However, transitions by other channels are possible: (1) multipolar emission, in which an emitter decays by changing its orbital angular momentum by more than one unit, and (2) multiphoton spontaneous emission (Fig. 4b), where an emitter decays by the *simultaneous* emission of multiple photonic quasiparticles. The rate of both types of processes is significantly enhanced by photonic quasiparticles in nano-cavities or polaritonic systems, because the field distributions of the quasiparticles becomes highly confined, such that the size of the electromagnetic field more closely matches the size of the wavefunction of the emitter.

Effects associated with multipolar transitions effects have been studied in the past using (metal) plasmonic nanoparticles both theoretically [Zurita-Sanchez2002a-b, Kern2012, Jain2013] and experimentally [Andersen2011], with some experiments demonstrating deviations from the classic dipole selection rules in metallic structures [Takase2013]. These beyond-dipole corrections were enhanced by the large electronic wavefunctions of the emitters used, namely carbon nanotubes [Takase2013] and mesoscopic quantum dots [Andersen2011]. In theoretical works, the focus was traditionally on electric quadrupole and magnetic dipole emission, the leading order beyond-dipole transitions, as higher-order decays were still weak relative to typical dipole transition rates. In comparison, effects associated with *simultaneous* emission of multiple photonic quasiparticles have only been studied in one experiment, which showed an enhanced two-plasmon emission in nanogap structures [Nevet2010].

Recently, it was predicted (in a unified manner via MQED) that polaritons in van der Waals materials can enable effectively-forbidden transitions due to their high confinement and local density of states. These transitions include high-order electric multipole transitions, singlet-triplet transitions, and even multiplasmon spontaneous emission (Fig. 4b) – all at rates approaching those of dipolar transitions in free space [Rivera2016]. Similarly, photonic quasiparticles (specifically graphene plasmons) were predicted to enable significant beyond-dipole effects in solid-state emitters such as quantum wells (Fig. 4c) – the emitter can absorb and emit light according to a *non-vertical* transition, thus changing its momentum significantly [Kurman2018]. The resulting non-vertical transitions lead to Doppler shifts, and are a manifestation of an induced spatial nonlocality in the quantum well.

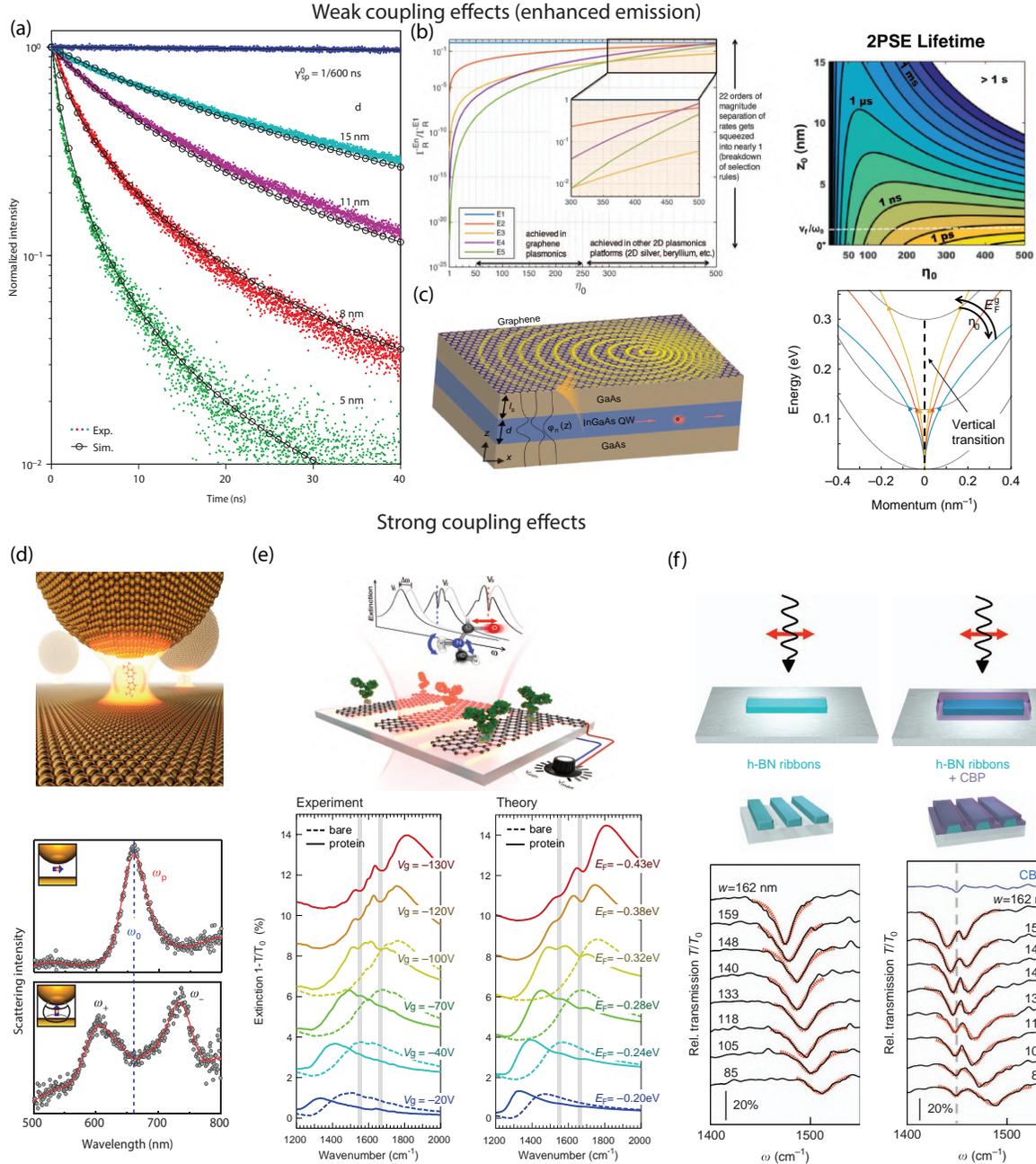

**Figure 4: Bound-electron interactions with photonic quasiparticles.** (a) In the visible spectral range, a molecular dye emitter in a plasmonic nanogap can have its spontaneous emission enhanced by nearly four orders of magnitude, reaching picosecond timescales (probed by time-resolved fluorescence) [Akselrod2014]. (b) Proposal to use highly confined plasmons to strongly enhance dipole-forbidden transitions and multi-photon emission processes. The strong confinement allows forbidden transitions to compete with conventionally allowed transitions, as well as allows two-plasmon emission processes to become comparable to one-plasmon processes [Rivera2016]. (c) The high momentum of a graphene plasmon allows significant momentum transfer from the electromagnetic field to electrons in a quantum well. Such a realization of optical nonlocality strongly changes absorption and emission spectral peaks [Kurman2018]. (d) When the light–matter coupling is strong enough, as in extremely small plasmonic nanogap cavities, even a small number of emitters can reach the strong-coupling regime, leading to Rabi splitting in the scattering spectrum [Chikkaraddy2016]. Strong coupling can also be realized by coupling many emitters (e.g., molecules) to a tightly-confined polariton mode, which can be used for (e) sensing molecules, as demonstrated with graphene plasmons [Rodrigo2015], and (f) infrared spectroscopy, as demonstrated with boron nitride resonators [Autore2018].

Going beyond the above predictions, recent theoretical works have proposed using phonon polaritons to make two-phonon-polariton emission dominate the single-phonon-polariton decay that enables strong quantum nonlinearities [Rivera2017], using plasmons with orbital angular momentum to control optical selection rules [Machado2018, Konzelmann2019], reaching strong coupling effects in multipolar decay [Neuman2018, Cuartero-Gonzales2018], showing interference effects between different multipolar channels [Rusak2019], using surface magnon polaritons to strongly enhance spin relaxation [Sloan2019], and reaching effects of spatial non-locality on multipolar and multi-plasmon transition enhancement in metals [Goncalves2020]. The last work shows applications of MQED to non-local media [Scheel2012]. Meanwhile, recent experimental works have investigated selection-rule breakdown based on the polarization of plasmons [Zhou2017], nonlocal (finite-wavevector) effects in absorption of light by van der Waals quantum wells [Schmidt2018] and by graphene [Zhang2019], enhancement of quadrupolar transitions with surface plasmons in atomic gases [Chan2019], and enhancement of singlet-triplet decays with hyperbolic metamaterials [Roth2019].

### *2.2 Strong coupling effects with tightly confined photonic quasiparticles.*

*Interaction of photonic quasiparticles with a two-level system: the Rabi Hamiltonian.* When emission and absorption are sufficiently enhanced, an emitter is capable of coherently emitting and re-absorbing a photonic quasiparticle before it is lost (e.g., to radiative or dissipative losses) [Torma2014, Flick2018, Forn-Diaz2019, Kockum2019]. The emitter and the cavity are then said to be in the *strong coupling* regime. A simple description of the strong coupling can be derived from the fundamental MQED Hamiltonian in the case where the emitter is strongly coupled to one mode, which is nearly resonant with a transition between two particular levels in the system (see Box 2). In that case, the MQED description becomes equivalent to the Rabi Hamiltonian $H_R$, i.e., a two-level system coupled to a single harmonic oscillator (the cavity mode):

$$H_R = \frac{\hbar\omega_0}{2}\sigma_z + \hbar\omega a^\dagger a + \hbar g \sigma_x(a + a^\dagger), \text{ with } g = \sqrt{\frac{\omega}{2\epsilon_0 \hbar V}} \boldsymbol{d}_{fi} \cdot \boldsymbol{u}(\boldsymbol{r}), \qquad (5)$$

where $\omega_0$ is the emitter frequency, $\omega$ is the cavity frequency, $\sigma_{z,x}$ are Pauli z- and x-matrices, and $a^{(\dagger)}$ is the annihilation (creation) operator for the cavity photon. The Rabi frequency $g$, which measures the strength of the interaction between matter and photon, can be found through MQED at different levels of approximation (see Box 2). In the case of a low-loss cavity and a dipole emitter at point $\boldsymbol{r}$, $g$ can be expressed in terms of the dipole moment of the transition $\boldsymbol{d}_{fi}$, the mode volume $V$, and the mode function $\boldsymbol{u}(\boldsymbol{r})$. It measures the interaction energy of the dipole with the vacuum(-fluctuation) field of the cavity.

One of the key phenomena encoded in this Hamiltonian is *Rabi splitting*. In particular, if the emitter and cavity are resonant with each other, and $g \ll \omega$, then the first two excited states of the system split in energy by an amount $2g$. This Rabi splitting is the hallmark of strong coupling phenomena, and is a key measurement in many works presenting evidence for strong coupling. Typically, this measurement proceeds by sending light at the strongly coupled system, and recording a scattering (e.g., transmission) spectrum. In the strongly coupled regime, the scattering spectrum will feature two resonances, split by the Rabi splitting (in contrast to a single resonance in the weakly coupled system). The strong coupling regime is manifested experimentally when the splitting is resolvable compared to the widths (related to losses). This condition is mathematically expressed as: $g > \sqrt{\kappa^2 + \gamma^2}$, with $\kappa$ the photonic quasiparticle loss, and $\gamma$ the atomic loss. The temporal dynamics associated with this frequency splitting are *damped vacuum Rabi* oscillations, in which the emitter coherently emits and re-absorbs the photonic quasiparticle multiple times before the quasiparticle decays. Such dynamics have been observed many times in the context of low-loss dielectric cavities [Wallraff2004, Reithmaier2004, Birnbaum2005, Aoki2006], but only recently have been observed in plasmonic contexts [Chikkaraddy2016, Santhosh2016].

In general, there are three ways to achieve strong coupling: by having many ($N$) emitters couple to the same mode ($g$ becomes enhanced by $\sqrt{N}$), by having many ($n$) photons pre-populate the cavity mode ($g$ becomes enhanced by $\sqrt{n}$), or by having a single emitter couple to an extremely confined mode with a small mode volume (since $g \sim 1/\sqrt{V}$). The last option represents strong quantum electrodynamical interaction at the single photon level. Any combination of these three methods enhances the coupling further. From the standpoint of this review, the strong confinement of the photonic quasiparticles considered here (polaritons, as well as highly-confined gap plasmons) can enable strong coupling with relatively few emitters and potentially, even a single emitter.

Fig. 4d illustrates this last point, showing a recent experiment demonstrating strong coupling of molecules to a nanoparticle-on-mirror geometry [Chikkaraddy2016]. The authors rely on a nanoparticle-on-mirror-geometry, based on a 0.9 nm gap established by a molecular spacer layer (cucurbit[7]uril) between a gold nanoparticle and a gold film. Beyond the use of cucurbit[7]uril as a spacer, it also intriguingly acts as a "cage" for the emitter used in the experiments, methylene blue, which also allows it to bind to the nanoparticles above. This gap structure supports extremely confined gap modes, which such a small mode volume, that the authors predict that the associated Purcell factors are of $3 \times 10^6$. These extreme enhancements are sufficient enough for a few emitters (between one and ten) to experience strong coupling to the cavity mode, as shown through measurements of the Rabi splitting as a function of the relative concentration of the emitter and the cucurbit[7]uril host. Since the density of molecules change the resonance frequencies of the combined system, the Rabi splitting can act as a measure of their concentration, allowing sensing applications.

Similar Rabi splittings can be observed in scattering spectra even in systems with less drastic (but still large) confinement by means of coupling more emitters to the mode [Endnote2]. Examples of this are shown in Fig. 4(e,f), specifically for 2D material systems: a graphene plasmon-based (bio)sensor [Rodrigo2015], and a hexagonal boron nitride phonon-polariton based sensor [Autore2018]. In both of these examples, it is the very strong (~10 nm scale) field confinement of the polariton, in conjunction with having many emitters, that enables strong coupling.

*Towards ultrastrong coupling.* As an outlook on this section, we mention one last very interesting theoretical possibility, related to *single-emitter ultrastrong coupling*, that can be achieved as the confinement of the photonic quasiparticle becomes such that it is comparable to the scale of the electronic wavefunction. Strong coupling, as discussed in the previous paragraphs, is maintained when the emitter's decay exceeds the loss rate, which is typically much smaller than the mode frequency. However, a new regime of quantum light–matter interactions emerges when the decay exceeds the mode frequency [Kockum2019, Forn-Diaz2019].

In that case, a number of phenomena emerge that do not occur in strong coupling. Interesting examples include: (1) Rabi oscillation even when an emitter is interacting with a continuum of modes (as in a waveguide, as opposed to a discrete cavity mode) [Forn-Diaz2017]. (2) Considerable changes in the energies of the *ground state*, due to very strong Lamb shift [Yoshihara2017, Forn-Diaz2017], which could allow changes in macroscopic thermodynamic properties such as chemical reactivity, specific heat, and even dielectric properties. (3) Virtual photons appear as part of the ground state of the coupled system (nonzero expectation values of photon number), which can in principle can be extracted by time-modulating the system, as in the *dynamical Casimir effect* [Ciuti2005, Wilson2009]. (4) Decoupling of light and matter for extreme coupling strengths [Liberato2014, Garcia-Ripoll2015, Rivera2019b]. The origin of many of these striking new phenomena is the breakdown of the rotating-wave approximation, in which one neglects the effect of virtual (energy non-conserving) processes, such as an emitter both becoming excited and emitting a photon, or an emitter becoming de-excited and absorbing a photon.

To this date, single-emitter ultrastrong coupling has only been observed in systems of superconducting qubits coupled to microwave cavities [Niemczyk2010, Forn-Díaz2010], which works due to the extremely large effective dipole moment of the qubit (which $g$ is proportional to). Looking forward, extremely confined graphene plasmons (as in [Iranzo2018]) may enable bringing single-emitter ultrastrong coupling to the infrared regime, as first suggested in [Rivera2016] and predicted theoretically to be possible in a graphene–quantum-well stack [Kurman2020].

**3. Light–matter interactions with photonic quasiparticles: free electrons**

*3.1 Controlling free electron spontaneous emission with photonic quasiparticles.*

Much of the focus in the field of quantum light–matter interaction is focused on emission and absorption of photonic quasiparticles based on *bound electrons*, i.e., emitters which are spatially confined by some potential in at least one dimension leading to discrete states or bands. However, many researchers are now considering the classical and quantum interactions of emitters based on *free electrons*. Part of the uniqueness of light–matter interactions of free electrons arises from their energy spectrum being continuous, rather than discrete as with most bound electron systems. This difference results in free-electrons transitions and free-electron radiation sources being tunable. Moreover, free electrons reach much higher (often relativistic) energies, which consequently enables transitions at much higher frequencies than is available for bound electron systems, even allowing emission of X-rays. In this section, we will go into detail on light–matter interactions enabled by free-electrons.

*Spontaneous emission by a free electron in a homogeneous medium: the Cherenkov effect.* We start by considering the Cherenkov effect, as in some sense, it represents the most basic light–matter interaction possible in free-electron systems. Indeed, the Cherenkov effect can be described as spontaneous emission by a free electron [Ginzburg1940, Sokolov1940]. Historically, the Cherenkov effect (or Cherenkov radiation) has been associated with the radiation emitted when a charged particle (not limited to free electrons) moves faster than the phase velocity of light in a homogeneous dielectric medium [Cherenkov1934]. Famously, in a non-dispersive medium, the radiation is emitted into a forward propagating cone centered around the direction of motion of the particle, with an opening angle $\theta$ that satisfies $\cos\theta = 1/\beta n$. Here $n$ is the index of refraction of the medium and $\beta = v/c$ is the speed of the particle ($v$) normalized to the speed of light ($c$) [Tamm1937]. Here, the effect is enabled because the photonic quasiparticle, i.e. the photon in a medium, has a phase velocity $v_p$ slower than $c$.

The scope of the Cherenkov effect goes far beyond charged particles in homogeneous media. For example, consider the relation $\cos\theta = 1/\beta n$. This relation is a specific way of representing a more general phase-matching condition that applies to many free-electron radiation processes beyond the Cherenkov effect. This condition is given by $\boldsymbol{v} \cdot \boldsymbol{k} = \omega(\boldsymbol{k})$, where $\boldsymbol{v}$ is the charged particle velocity, $\boldsymbol{k}$ is the wavevector of the photon emitted, and $\omega(\boldsymbol{k})$ the corresponding frequency of the photon prescribed by the dispersion relation [Friedman1988, Luo2003, DeAbajo2010, Gover2019a]. This phase-matching condition is a result of energy and momentum conservation. Moreover, as can be seen from this general phase-matching, the emitted photon need not be in a homogeneous medium. The emission can be into a more general photonic quasiparticle, such as a waveguide mode, or surface polaritons such as plasmon and phonon polaritons, or a photonic Bloch mode, provided that the system has a well-defined momentum in some direction (discrete or continuous translation symmetry). It is also possible for the electron to emit into localized (cavity) modes, analogously to much of the research investigating bound electron coupling to cavities.

Previous work showed how the dispersion relation of the photon distinguishes between variants of the Cherenkov effect. For example, negative index materials [Veselago1967], and photonic crystals [Luo2003], both were predicted to support a backward Cherenkov cone. This effect was observed using mathematical analogies simulating the emitting particle by a phased-array antenna [Xi2009]. Mathematical analogies were also used to observe a kind of Cherenkov effect involving a directional emission of surface plasmon polaritons, using metasurfaces [Genevet2015] to simulate the polarization field of a moving electron. Nevertheless, such effects have yet to be observed with true charged particles. Controlling the angular emission properties of Cherenkov radiation is important, particularly in applications such as Cherenkov-based particle detectors, where it is the emission angles that are used to determine the properties of incident high-energy particles [Lin2018].

These interactions can be measured by detecting the electron energy losses associated with the emission (via electron-energy loss spectroscopy EELS) [deAbajo2010], which was also shown to probe the local density of states of the optical structure. That the electrons probe the local density of states can be seen from the general expression for spontaneous emission by a quantum system of Eq. (2). The total rate $d\Gamma_i$ of energy loss by an electron in an initial energy eigenstate $i$ (e.g., a plane wave with some momentum $\hbar\boldsymbol{k}$), into all possible final states, per unit energy-loss $d\omega$, is given by:

$$\frac{d\Gamma_i}{d\omega} = \frac{2\mu_0}{\hbar}\sum_f \int d^3r\, d^3r'\, \boldsymbol{j}_{fi}^*(\boldsymbol{r}) \cdot \text{Im}\, \boldsymbol{G}(\boldsymbol{r},\boldsymbol{r}',\omega) \cdot \boldsymbol{j}_{fi}(\boldsymbol{r}')\delta(\omega - \omega_{if}). \tag{6}$$

Note that because of the extended nature of a generic free electron, the electron probes a more general quantity than the local density of states, as the electron probes the Green's function at two *different* locations. However, it is often the case, as in high-resolution electron microscopes, that the emitter is an electron wavepacket which is well-localized around a straight-line trajectory

$\boldsymbol{r}(t) = \boldsymbol{r}_0 + \boldsymbol{v}t$. It can then be seen by direct application of Fermi's Golden Rule that the probability $d\mathrm{P}$ of the electron of losing energy $\hbar\omega$ per unit frequency $d\omega$ is given by

$$\frac{d\mathrm{P}}{d\omega} = \frac{\mu_0 q^2}{\pi\hbar} \int dt\, dt'\, e^{i\omega(t-t')} (\boldsymbol{v} \cdot \mathrm{Im}\, \boldsymbol{G}(\boldsymbol{r}_0 + \boldsymbol{v}t, \boldsymbol{r}_0 + \boldsymbol{v}t', \omega) \cdot \boldsymbol{v}), \qquad (7)$$

which is the standard EELS formula [deAbajo2010]. Eq. (7) thus shows that the electron probes the local density of states along its trajectory, for an arbitrary optical structure [deAbajo2010, Sapienza2012, Peng2019].

*The underlying quantum nature of the Cherenkov effect*. The vast majority of the research done on the Cherenkov effect has been based purely on classical electrodynamics, which has accounted perfectly for all known observations thus far. The Cherenkov effect can also be explained through MQED [Ginzburg1940, Sokolov1940] (Fig. 3) simply as the equivalent of spontaneous emission by a free-electron in a medium. This equivalence emphasizes the central place of the Cherenkov effect in the light–matter interactions of free electrons. Moreover, the quantum treatment of Cherenkov radiation leads to corrections originating from the recoil of the emitting particle due to the emission of a single quanta of photonic quasiparticle. The quantum recoil corrections have been predicted to be significant in certain conditions for the Cherenkov effect in regular materials [Kaminer2016b] and in graphene [Kaminer2016a], and for low energy electrons in the analogous Smith-Purcell effect [Tsesses2017].

Another type of a quantum correction exists in the Cherenkov effect and in other electron radiation phenomena (Fig. 5): the dependence of radiation emission on the wavefunction of the emitting particle. Such phenomena have been predicted for the Cherenkov effect [Kaminer2016b], Smith-Purcell effect [Remez2019], other spontaneous radiation mechanisms [Murdia2017], and their stimulated analogues [Gover2019b]. The first few experiments on this effect have been performed in recent years. One experiment showed no wavefunction dependence [Remez2019] because the emission did not depend on characteristics of the photonic quasiparticle, and could be modeled with free-space photons. In contrast, an indirect measurement through EELS showed the first evidence of a wavefunction effect in the other extreme case, of emission into localized surface plasmons [Guzzinati2017], where the characteristics of the photonic quasiparticles deviated significantly from those of a free photon. The key difference in these experiments is the nature of the photonic quasiparticle. For the precise shape of the wavefunction to influence the radiation, it must be the case that: two electron states can transition to the same final electron and photon state, so that the transition amplitudes can interfere. In the case of Smith-Purcell radiation of a 1D grating, the photonic quasiparticle has a well-defined momentum (up to a lattice vector), and then strict momentum conservation does not allow two distinct electron states to interfere. In the case of localized quasiparticles (as in [Guzzinati2017]) that break translation invariance, such an interference becomes possible due to the relaxation of conservation laws. Therefore, in contrast to previous cases in this Review, where the dispersion, or confinement, or polarization was the root cause of the effects, here it is *symmetry*.

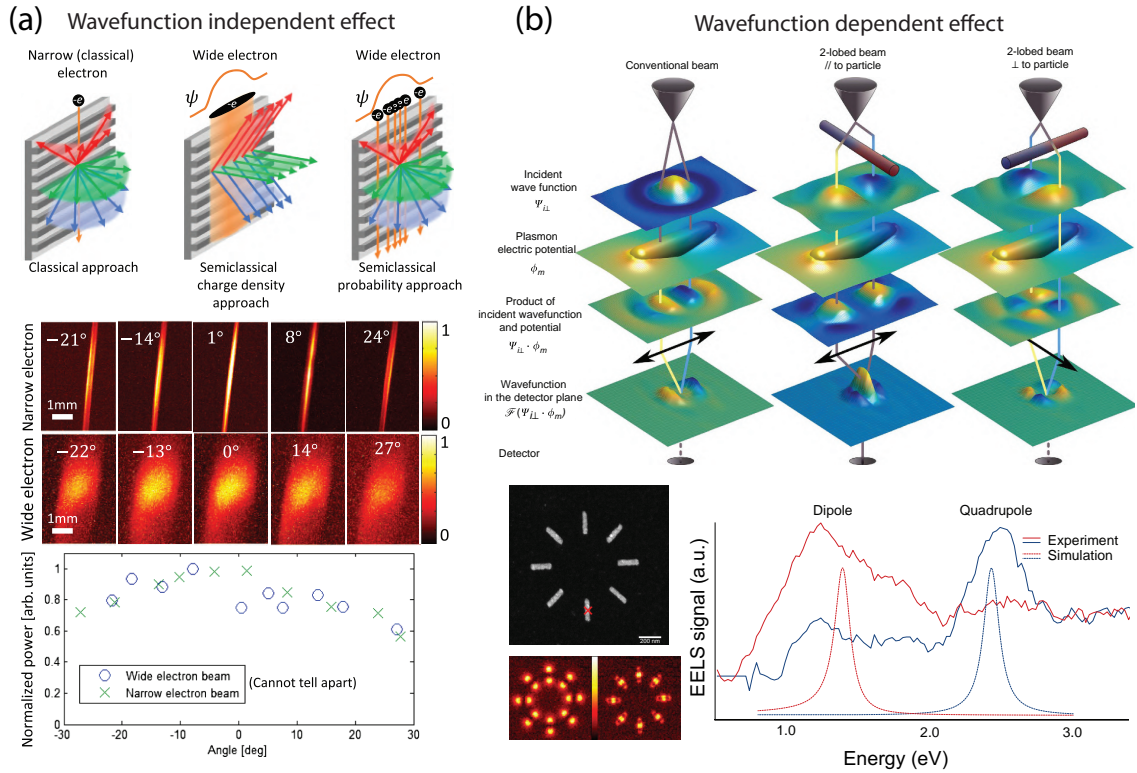

**Figure 5: Free-electron spontaneous radiation: quantum wavefunction-dependent effects.** (a) Experiment showing the effect on the coherent size of the electron wavefunction on Smith-Purcell radiation. Comparing a narrow versus wide electron wavefunction, a change is seen in the spatial distribution of the radiation (see color-maps) but no influence is seen on the radiated angular power spectrum [Remez2019]. (b) The influence of photonic quasiparticles on electron energy-loss spectroscopy, through symmetry-matching between plasmonic modes and the electron wavefunction, showing how one can control which plasmon modes are coupled to by shaping the electron wavefunction to have a matching symmetry [Guzzinati2017].

*The Cherenkov effect in condensed matter physics as a test of photonic quasiparticles*. Modern incarnations of the Cherenkov effect (Fig. 6) demonstrate the wide applicability of photonic quasiparticles. Specifically, we show that the Cherenkov effect has now been studied with plasmons (Fig. 6a) [deAbajo2013, Liu2014, Sundararaman2014, Brown2016, Kaminer2016a] and with phonons in solids (Fig. 6b) [Zhao2013, Andersen2019], which are the photonic quasiparticles that interact with ultra-slow electrons in solids (in place of relativistic electrons). The emission follows the same phase-matching condition $\bm{v} \cdot \bm{k} = \omega(\bm{k})$, up to quantum recoil corrections. Here, these effects are enabled with electrons in solids because the photonic quasiparticle, the bulk plasmon or phonon, has a phase velocity two-to-four orders of magnitude slower than $c$.

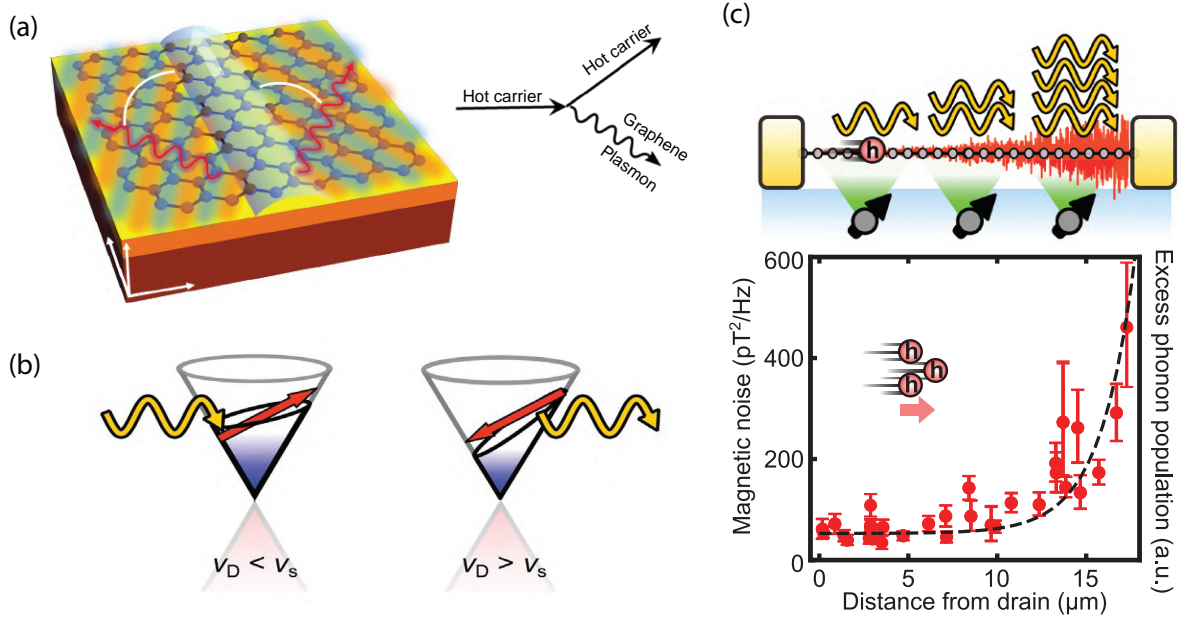

**Figure 6: The similarity of electron-photon, electron-plasmon, and electron-phonon interactions: all can be understood as interactions with photonic quasiparticles** (a) Cherenkov radiation of plasmons in graphene by hot electrons can occur with a very high efficiency compared to Cherenkov radiation in transparent dielectric media. (b) Cherenkov emission of phonons by electrons in ultraclean graphene [Kaminer2016a]. (c) Such Cherenkov emission of phonons has been used to explain the amplification of magnetic noise by electrons moving in ultraclean samples of graphene [Andersen2019].

Taking phonons specifically, their slow velocities enable electrons and holes in solids to emit phonons in a Cherenkov effect, as well as absorb them in an inverse Cherenkov effect. These phenomena can occur in conventional solids [Pippard1963, Hutson1961] and in graphene [Zhao2013]. Such Cherenkov processes are equivalent to charge carrier relaxation and thermalization by electron-phonon scattering. Nevertheless, treating the process through a Cherenkov formalism proved useful in explaining recently observed phenomena of electron-phonon instabilities and noise amplification in graphene [Andersen2019]. Beyond these effects with electrons in solids, relativistic free electrons are also used to probe phonons through measuring the energy losses of electrons that spontaneously emit phonons. Such techniques are now used for vibrational spectroscopy [Hage2019, Venkatraman2019, Hachtel2019]. Similar to phonon scattering, even charge carrier scattering (Landau damping) by surface and by bulk plasmons can be connected to a Cherenkov-like process as pointed out by Ginzburg [Ginzburg1996]. This similarity between all the excitations helps promote the combined treatment of all photonic quasiparticles with the same concepts and methods of light–matter interactions, as shown in Fig. 3. This combined treatment shows that despite the different microscopic origins of electromagnetic excitations, and despite their varying degree of photon vs. matter composition, they can all be considered as instances of a more general photonic quasiparticle.

*Spontaneous emission by a free electron in a periodic medium: the Smith-Purcell effect.* Being quite similar in essence to the Cherenkov effect, the Smith-Purcell effect has an electron traveling along a *periodic* optical system, and emitting light into the far-field [Smith1953]. The effect can be understood from the Cherenkov effect, but using a different photonic quasiparticle, which is the Bloch photon mode. Here, the effect is enabled because the photonic quasiparticle, the Bloch photon, has higher momentum components associated with additions of reciprocal lattice vectors. An electron can couple to a Bloch photon if $\mathbf{v} \cdot (\mathbf{k} + \mathbf{G}) = \omega_k$ (i.e., phase-matching) is satisfied, where $\mathbf{k}$ is the Bloch wavevector inside the first Brillouin zone, and $\mathbf{G}$ is a reciprocal lattice vector [Friedman1988, Abajo2010, Gover2019a]. Smith-Purcell radiation arises when this (evanescent) harmonic of wavevector $\mathbf{k} + \mathbf{G}$ diffracts into the far-field. The frequency of the emitted photon depends on the angle of emission, and the periodicity of the crystal by the famous relation $\omega = (\mathbf{v} \cdot \mathbf{G})/(1 - \beta \cos \theta)$, showing that emission into gratings with small periods enables high-frequency (even ultraviolet radiation [Ye2019]), motivating a push to observe Smith-Purcell effects (and other related free-electron radiation effects) in the interaction of free-electrons with nanostructures [Adamo2009, Kaminer2017, Clarke2018, Ye2019]. Smith-Purcell radiation is possible for any periodic medium, both metallic gratings where Smith-Purcell was historically studied (and explained in terms of image charges) [Smith1953, Doucas1992], and dielectric gratings, e.g., silicon [Roques-Carmes2019]. In all cases, by modeling Smith-Purcell radiation as the grating scattering (diffraction) of the electron's near-field into the far-field, one can derive fundamental bounds on the efficiency of Smith-Purcell radiation, as developed and probed experimentally in [Yang2018].

*Spontaneous emission by free electrons in strong fields of photonic quasiparticles.* As an outlook on the possibility of applying the considerations of Fig. 3 to free-electron processes, we discuss recent theoretical proposals related to free-electron radiation in strong driving fields. Both the strong driving field and the emitted radiation can be modified by the optical environment and lead to new effects. In particular, the electron can absorb or stimulatedly emit photonic quasiparticles from an external driving field *and* spontaneously emit another photonic quasiparticle. Typically, due to the relativistic nature of the emitting electron, the spontaneously emitted photon can be at a much different frequency from the original photon. As an example, Fig. 7a shows a proposal to scatter free electrons from a strongly pumped external plasmonic standing wave on the surface of graphene [Wong2016]. The free-electron can then undergo a Compton-like process in which it absorbs (or stimulatedly emits) the plasmon and emits a photon. Due to the relativistic nature of the electron *and* the high optical confinement of the plasmon, the emitted photon can be at hard X-ray frequencies. Compared to other sources of X-rays, this source can produce X-rays using much less relativistic electrons due to the graphene plasmon confinement. That said, the small extent of the evanescent graphene plasmon strongly limits the achievable flux/intensity, with heterostructures having been proposed as a method to mitigate this [Rosolen2018, Pizzi2020].

Interestingly, such radiation processes can in fact take place *without any driving field*. Fig. 7b considers the case in which a free-electron spontaneously emits *both* the plasmon and the X-ray photon, which is equivalent to Compton scattering from *plasmonic vacuum fluctuations* [Rivera2019a]. Strikingly, such a spontaneous process has similar power yields as the stimulated process due to the very strong vacuum fluctuations on the nanoscale, though the emission is far less monochromatic, due to the heavily multimode nature of the process (i.e., spontaneous emission occurs into any available plasmon mode, leading to X-ray emission at a wide spectrum).

So far, all the considered processes were first- or second-order in MQED, but there also exist radiation processes in which many photons are absorbed or stimulatedly emitted (effectively higher-order MQED), followed by spontaneous emission of a single photonic quasiparticle. Such

*nonlinear Compton scattering processes*, are typically very weak, but can become efficient when the emission is into *plasmons* due to their strong confinement [Rivera2019c]. This enhancement is a manifestation of the Purcell effect, but for strongly driven free electrons (Fig. 7c) instead of bound electrons.

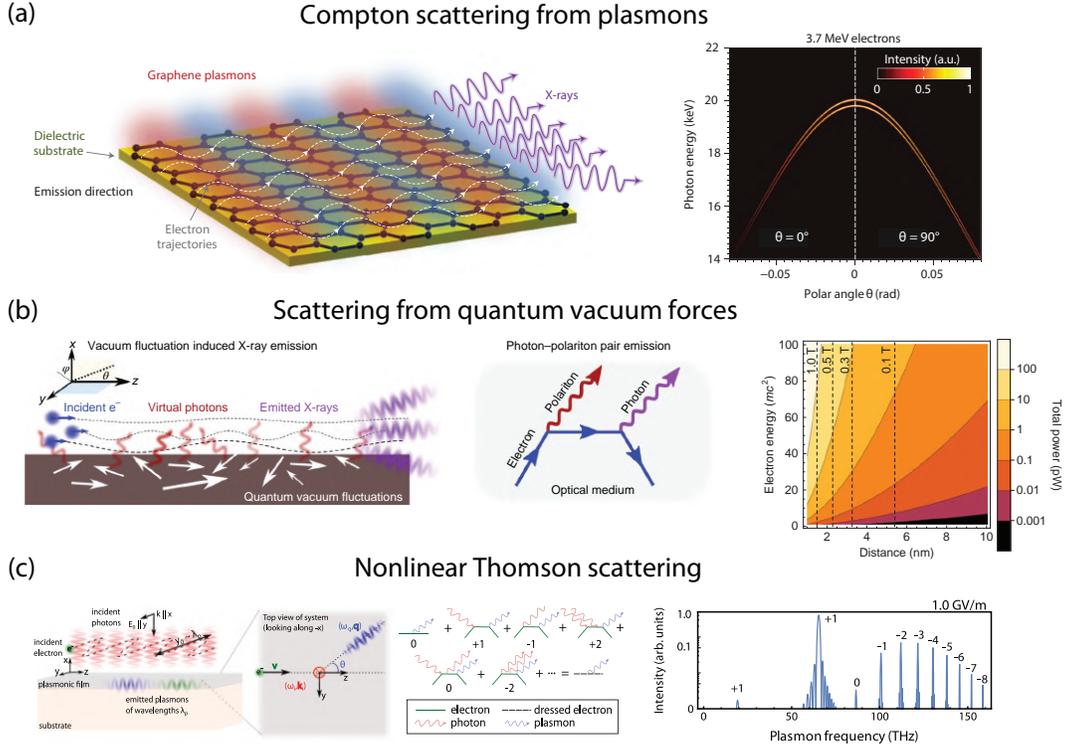

**Figure 7: Free-electron spontaneous radiation: the influence of photonic quasiparticles.** (a) Proposal of laser-driven photonic quasiparticles, in the form of surface plasmons, that produce X-rays from free electrons via inverse Compton scattering (up-converting the plasmon into X-rays) [Wong2016]. (b) The phenomenon can even occur without an externally excited plasmon, using strong Casimir-type forces based on vacuum fluctuations of photonic quasiparticles [Rivera2019a]. (c) Proposal to generate high harmonics of photonic quasiparticles, in the form of surface plasmons, by electrons interacting with strong fields. The emitted plasmonic pulses can reach nanometer and femtosecond-scale with a comb-like profile [Rivera2019c].

### *3.2 Strong coupling of free electrons and photonic quasiparticles.*

*Stimulated emission and absorption of photonic quasiparticles*. Photonic quasiparticles can be used to exert a great deal of control over spontaneous emission by free electrons, in a similar way as for bound electrons. A natural question, extending ideas from bound electron physics, is whether or not strong or ultrastrong coupling (and associated phenomena, such as Rabi oscillations) can also be realized in free-electron systems. Here, some distinction should be made between *vacuum* strong coupling effects, where the electron-light coupling $g$ is strong enough to induce Rabi oscillations, and *stimulated* strong coupling effects. In the case of stimulated effects, the coupling is effectively enhanced to $\sqrt{n+1}g$ in the presence of $n$ photonic quasiparticles (see Section 2.2). This enhancement is similar to the case in bound electron systems, where Rabi oscillations in atoms, molecules, and various types of qubits, can be induced by a strong driving field. Vacuum strong and ultrastrong coupling has not yet been observed with free electron systems, though there have been some proposals for strong coupling [Kfir2019] based on electron-cavity interactions. Other proposals for vacuum ultrastrong coupling involved Cherenkov radiation by heavy ions [Roques-Carmes2018] and Cherenkov radiation of graphene plasmons by electrons in solids [Kaminer2016a].

Strong coupling and ultra-strong coupling effects have been observed in non-relativistic systems of particles which are closely related to free electrons. In particular, strong coupling, and the associated phenomena of Rabi splitting in scattering spectra, have been observed in 2D electron gas systems (2DEGs) associated with high-mobility quantum wells immersed in magnetic fields. These systems feature many electrons occupying Landau levels which are collectively coupled to a common resonant cavity mode, typically a terahertz cavity mode associated with a metallic resonator hosting a highly-confined mode [Scalari2012, Bayer2017]. Because strong coupling modifies the energy spectra of the composite system, and because macroscopic properties such as transport and other thermodynamic properties depend on the underlying energy spectra of the system, strong coupling can change the intrinsic properties of the system. This was demonstrated very recently in the context of magnetotransport of electrons in 2DEGs, where the transport properties were strongly modified by the presence of a terahertz resonator [Paravicini-Bagliani2019]. Similar Rabi splitting effects have also been observed in the coupling of cavities to other free-electron-like systems, such as intersubband transitions in quantum wells (through their electric dipole moments) [Dini2003, Todorov2010], and even collective excitations of cooper-pairs (Josephson plasma resonances) [Schalch2019].

*Photon-induced near-field optical microscopy (PINEM).* While *vacuum* strong coupling effects were not observed so far with free electrons, *stimulated* strong coupling effects have emerged in recent years using pulses of free electrons interacting with pulses of strong laser fields [Barwick2009]. These results have had immediate applications in ultrafast electron microscopy [Zewail2010].

The most influential advances in this direction are the results of the new capability called PINEM [Barwick2009], in which an electron interacts with a strong field that is coupled to a material. The electron undergoes absorption and stimulated emission of many photons of the driving field in a way that also provides new insight on the material. This absorption and emission can be shown to be equivalent to a multi-level quantum system with equally spaced energy levels undergoing quantum Rabi oscillations [Feist2015]. The number of photons absorbed and emitted scales with a dimensionless parameter $g$. This parameter is also equivalent to a quantity used in linear-field laser-acceleration in accelerator physics [Peralta2013, England2014, Dahan2019, Sapra2020] – the integrated work done by the component of the electric field ($E_z$) along the trajectory of the particle of charge $q$, which in our case is normalized by the energy of the driving photon $\hbar\omega$. For an electron moving with velocity $v$ along the $z$ direction, $g$ equals

$$g = \frac{q}{\hbar\omega} \int_{-\infty}^{\infty} dz \, e^{-i\omega z/v} E_z(z). \tag{7}$$

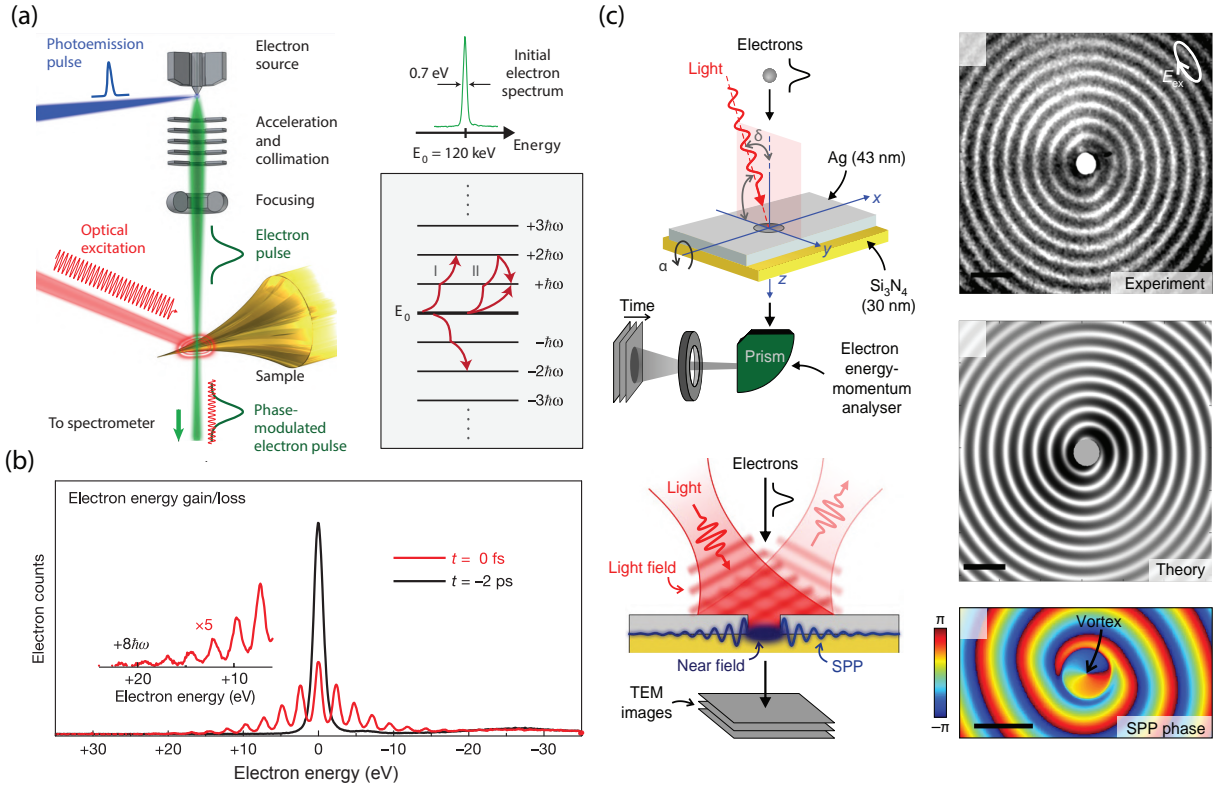

**Figure 8: Effects enabled by strong fields of photonic quasiparticles:** photon-induced near-field electron microscopy (PINEM). Stimulated electron-photon interactions when driving laser fields pump a photonic quasiparticle mode, as demonstrated in PINEM. Each electron undergoes stimulated absorption and emission of multiple photons as a result of the PINEM interaction with a strong field, leading to quantized energy gain and loss. (a) The electron can be seen as undergoing a quantum walk on the energy ladder with spacing set by the driving frequency [Feist2015]. (b) First demonstration of PINEM [Barwick2009]. (c) When the electron interacts with a chiral plasmonic field, it imparts orbital angular momentum to the electron, seen in its diffraction pattern [Vanacore2019].

The PINEM interaction has been observed for a wide-range of photonic quasiparticles, from: localized plasmons [Piazza2015], surface plasmons [Lummen2016], free space plane wave scattering off a mirror [Vanacore2018], photonic crystal modes [Wang2019] and whispering gallery modes [Kfir2019], as well as propagating photonic modes in a half-infinite homogeneous medium [Dahan2019]. In all cases, the presence of matter that modifies the free-space photon is critical, as the equation for $g$ vanishes for any field $E$ in free-space. This result shows the necessity of a strong driving laser pumping an electromagnetic field mode that deviate from that of free-space so that the integral of $g$ does not vanish.

The experimental setups used for such interactions are ultrafast transmission electron microscopes [Zewail2010], with related effects also observed in ultrafast electron diffraction setups [Priebe2017, Morimoto2018] and in other electron-beam setups [Kozak2018], which show the classical corresponding effects of PINEM. Fig. 8 presents exemplary experimental results in the field, including the extremely nonlinear interaction of a free electron with multiple photons (i.e, ten-photon absorption/stimulated emission Fig. 8b [Barwick2009]) creating free-electron Rabi oscillations Fig. 8a [Feist2015]. This nonlinear interaction has been applied in microscopy for imaging plasmons at buried interfaces [Lummen2016], presenting meV energy resolution in EELS

[Pomarico2017], and imaging plasmons with angular momentum (Fig. 8c) [Vanacore2019]. The latter uses the quantized nonlinear interaction of electrons with the angular-momentum carrying plasmons to create electron vortex beams [Bliokh2007].

### 4. Outlook

In this Review, we have surveyed the broad physics of the interactions between bound/free electron emitters and photonic quasiparticles (photons in media). We showed that by using the photonic quasiparticle concept (Figs. 1, 2) to describe any electromagnetic field in a medium, we could understand many seemingly disparate phenomena (Fig. 3) by appealing to either the confinement (Figs. 4 and 7), symmetry (Figs. 5 and 8), or dispersion (Figs. 6, 7, 8) of the photonic quasiparticle.

We emphasize here that the photonic quasiparticle is rigorously supported by MQED, which allows one to quantize electromagnetic fields in *any* medium, including non-local ones. One can quantize photons in vacuum, in transparent media, cavity photons, Bloch photons, polaritons in van der Waals materials, and even bulk phonons and plasmons (which are described by non-local response functions). MQED thus serves as a key unifying tool in the physics of light–matter interactions.

From the point of view of MQED fundamentals, many opportunities still remain to be explored in light–matter interactions with photonic quasiparticles. We highlight some of the most ambitious directions here. Many open questions remain on the nature of ultrastrong coupling of emitters to systems with a continuum of modes. Can ultrastrong coupling be used to design new bound states of emitters with photonic quasiparticles? How can strong multiphoton effects be used to design materials with *stronger optical nonlinearities*? Another interesting direction regards the fact that energy levels of emitters can shift due to virtual absorption and re-emission of photonic quasiparticles, according to the Lamb shift. Can emitters be re-designed at will using Lamb shifts in the ultrastrong coupling regime? Such questions also beget questions regarding renormalization in MQED [Endnote3]. Finally, as an outlook on novel X-ray generation mechanisms, it is of practical interest to explore how/whether these mechanisms can serve as an effective *gain* medium at X-ray frequencies.

We emphasize here that this field is still in a nascent stage. There are still many theoretical directions to explore, and there are many predictions still waiting to be verified. More than half the experiments in PINEM have been published just in the past few years. Looking forward, it will be of interest to experimentally demonstrate spontaneous (Cherenkov-type) and stimulated (PINEM-type) interactions of free electrons with novel polaritons. In particular, the stimulated PINEM interactions may enable new methods to image the dynamics of highly-confined polaritons with nanometer and femtosecond resolution.

The most recent predictions on strong light–matter interactions with highly confined photonic quasiparticles in 2D materials have not yet been demonstrated experimentally. Thus, one of the most important goals moving forward will to be to test the exciting predictions made regarding enhancing spontaneous emission, realizing forbidden transitions, and achieving strong and even ultrastrong coupling phenomena in new material platforms at optical frequencies. Moreover, it has yet to be shown that enormous spontaneous emission enhancements also extend to two-photon processes. Another exciting experimental direction that we expect to see in the next few years is probing light–matter interactions of bound and free electrons with photonic quasiparticles in Moire systems [Ni2015, Sunku2018]. Such an experiment will eventually enable to observe strong coupling between twisted bilayer systems and optical cavities, altering the energy spectra of the Moire system, potentially influencing for example their transport and other macroscopic properties.

**Acknowledgements**

We would also like to acknowledge Yaniv Kurman, Alexey Gorlach, Ori Eyal, Jamison Sloan, Thomas Christensen, Prof. Dmitri Basov, Prof. Stefan Scheel, and Prof. Moti Segev for their helpful comments on the Review. We would also like to acknowledge Prof. Marin Soljacic and Prof. John Joannopoulos for the fruitful collaborations that led to this Review. Nicholas Rivera was supported by Department of Energy Fellowship DE-FG02-97ER25308, and a Dean's Fellowship by the MIT School of Science. Ido Kaminer was supported by the Azrieli Faculty Fellowship, the ERC starting grant NanoEP 851780 from the European Research Council, the Israel Science Foundation grant no. 831/19, and the GIF Young Scientists' Program by the German-Israeli Foundation for Scientific Research and Development.

**Box 1: Macroscopic quantum electrodynamics (MQED)**

| Photons in vacuum | |
|---|---|
| $\mathbf{A}(\mathbf{r}) = \sum_k \sqrt{\frac{\hbar}{2\epsilon_0 \omega_k V}} \left( e^{i\mathbf{k}\cdot\mathbf{r}} \hat{\epsilon}_k a_k + e^{-i\mathbf{k}\cdot\mathbf{r}} \hat{\epsilon}_k^* a_k^\dagger \right)$ | $\omega_k = ck$ <br> $\mathbf{k} \cdot \hat{\epsilon}_k = 0$ |

| Photons in a homogeneous medium | |
|---|---|
| $\mathbf{A}(\mathbf{r}) = \sum_k \sqrt{\frac{\hbar}{2\epsilon_0 \omega_k n^2 V}} \left( e^{i\mathbf{k}\cdot\mathbf{r}} \hat{\epsilon}_k a_k + e^{-i\mathbf{k}\cdot\mathbf{r}} \hat{\epsilon}_k^* a_k^\dagger \right)$ | $\omega_k = ck/n$ <br> $\mathbf{k} \cdot \hat{\epsilon}_k = 0$ |

| Photons in an inhomogeneous, lossless, non-dispersive medium | |
|---|---|
| $\mathbf{A}(\mathbf{r}) = \sum_n \sqrt{\frac{\hbar}{2\epsilon_0 \omega_n}} \left( \mathbf{F}_n(\mathbf{r}) a_n + \mathbf{F}_n^*(\mathbf{r}) a_n^\dagger \right)$ | $\nabla \times \nabla \times \mathbf{F}_n = \epsilon(\mathbf{r}) \frac{\omega_n^2}{c^2} \mathbf{F}_n$ <br> $\int d^3r\, \epsilon(\mathbf{r}) |\mathbf{F}_n|^2 = 1$ |

| Photons in an inhomogeneous, lossless, dispersive, anisotropic, medium | |
|---|---|
| $\mathbf{A}(\mathbf{r}) = \sum_n \sqrt{\frac{\hbar}{2\epsilon_0 \omega_n}} \left( \mathbf{F}_n(\mathbf{r}) a_n + \mathbf{F}_n^*(\mathbf{r}) a_n^\dagger \right)$ | $\nabla \times \nabla \times \mathbf{F}_n = \epsilon(\mathbf{r}, \omega_n) \frac{\omega_n^2}{c^2} \mathbf{F}_n$ <br> $\frac{1}{2} \int d^3r\, \mathbf{F}_n^* \cdot \frac{d\epsilon(\mathbf{r}, \omega)}{d\omega} \cdot \mathbf{F}_n = 1$ |

| Photons in an inhomogeneous, lossy, local medium | |
|---|---|
| $\mathbf{A}(\mathbf{r}) = \sqrt{\frac{\hbar}{\pi\epsilon_0}} \int_0^\infty d\omega\, \frac{\omega}{c^2} \int d^3r'\, \sqrt{\mathrm{Im}\,\epsilon(\mathbf{r},\omega)} \left( \mathbf{G}(\mathbf{r},\mathbf{r}',\omega) \mathbf{f}(\mathbf{r}',\omega) + \mathbf{G}^*(\mathbf{r},\mathbf{r}',\omega) \mathbf{f}^\dagger(\mathbf{r}',\omega) \right)$ | $\left( \nabla \times \nabla \times -\epsilon(\mathbf{r},\omega) \frac{\omega^2}{c^2} \right) \mathbf{G}(\mathbf{r},\mathbf{r}',\omega)$ <br> $= \delta(\mathbf{r}-\mathbf{r}') I$ |

| Photons in an inhomogeneous, lossy, nonlocal medium | |
|---|---|
| $\mathbf{A}(\mathbf{r}) = \sqrt{\frac{\hbar}{\pi\epsilon_0}} \int_0^\infty d\omega\, \frac{\omega}{c^2} \int d^3r'\, \sqrt{\mathrm{Im}\,\epsilon(\mathbf{r},\mathbf{r}',\omega)} \left( \mathbf{G}(\mathbf{r},\mathbf{r}',\omega) \mathbf{f}(\mathbf{r}',\omega) + \mathbf{G}^*(\mathbf{r},\mathbf{r}',\omega) \mathbf{f}^\dagger(\mathbf{r}',\omega) \right)$ | $\left( \nabla \times \nabla \times - \int d^3r'\, \epsilon(\mathbf{r},\mathbf{r}',\omega) \frac{\omega^2}{c^2} \right) \mathbf{G}(\mathbf{r},\mathbf{r}',\omega)$ <br> $= \delta(\mathbf{r}-\mathbf{r}') I$ |

**Table 1: Levels of quantization of the electromagnetic field**, showing the quantized vector potential operator under different cases of linear media, starting from the well-known field quantization in vacuum to quantization of the electromagnetic field in a truly arbitrary medium that can be inhomogeneous, anisotropic, lossy, and even spatially non-local.

In any general material, including lossy ones, we may represent an electromagnetic field operator in terms of a "mode expansion" that decomposes the electromagnetic field in terms of the fields of time-harmonic point dipoles in the medium. These dipoles are parameterized by their location $\mathbf{r}$, frequency $\omega$, and orientation $k = 1,2,3$ (or $x, y, z$). The quantization of the electromagnetic field proceeds by quantizing these dipoles, associating with each $(\mathbf{r}\omega k)$ a quantum harmonic oscillator with associated creation $f_k^\dagger(\mathbf{r}, \omega)$ and annihilation $f_k(\mathbf{r}, \omega)$ operators, satisfying $[f_k(\mathbf{r}, \omega), f_{k'}(\mathbf{r}', \omega')]$, $[f_k(\mathbf{r}, \omega), f_{k'}(\mathbf{r}', \omega')]^\dagger = 0$ and $[f_k(\mathbf{r}, \omega), f_{k'}^\dagger(\mathbf{r}', \omega')] = \delta_{kk'} \delta(\omega - \omega') \delta(\mathbf{r} - \mathbf{r}')$. Using these operators, the EM field Hamiltonian is given by

$$H_{\mathrm{em}} = \int_0^\infty d\omega \int d^3r\, \hbar\omega\, \boldsymbol{f}^\dagger(\boldsymbol{r}, \omega) \cdot \boldsymbol{f}(\boldsymbol{r}, \omega),$$

where we have left out the zero-point energy.

The resulting vector potential takes the form:

$$A(r) = \sqrt{\frac{\hbar}{\pi\epsilon_0}} \int d\omega \frac{\omega}{c^2} \int d^3r' \sqrt{\operatorname{Im}\epsilon(r',\omega)}(G(r,r',\omega) \cdot f(r',\omega) + G^*(r,r',\omega) \cdot f^\dagger(r',\omega)),$$

where $G(r,r',\omega)$ is the Green's function of the Maxwell equations, which in a non-magnetic medium satisfies $(\nabla \times \nabla \times - \epsilon(r,\omega)k^2)G(r,r',\omega) = \delta(r-r')I$, with $k = \omega/c$. $\epsilon(r,\omega)$ is the permittivity tensor in a general dispersive, local, anisotropic medium, and $I$ the $3 \times 3$ identity matrix. The MQED vector potential for the nonlocal case is shown in Table I. This quantized field operator is a central result of MQED [Scheel2008], and all the previous expressions for the quantized fields in terms of mode expansions are special cases of this. Note that all of the cases represented in Table 1 assume *non-magnetic* media. For magnetically polarizable media, as reviewed in [Scheel2008], additional $f$ operators must be introduced that correspond to magnetic dipole excitations, which are connected to the field operators through a magnetic Green's function. Then, the field operators are a sum of terms from electric and magnetic dipoles.

We briefly comment on the physical principles encoded in this formalism. The quantized field is connected to quantized dipoles through the *classical* Maxwell equations. We give a brief heuristic sketch of how these dipoles are quantized. For simplicity, we will do it here in an isotropic, local medium (which can still be lossy). The idea is to write a current field operator as a sum over bosonic degrees of freedom (point dipoles governed by position, frequency, and direction): $j(r) = \int_0^\infty \frac{d\omega}{2\pi} \int d^3r \, (N(r,\omega) f(r,\omega) + N^*(r,\omega) f^\dagger(r,\omega))$, with $N(r,\omega)$ some unknown normalization constant. The normalization is prescribed by both the commutation relations between the $f$s and the fact that the correlation functions must be in agreement with the fluctuation-dissipation theorem for a linear medium. In particular, for a linear medium, it must be the case that $\langle j(r,\omega) \otimes j(r',\omega) \rangle = \epsilon_0 \hbar \omega^2 \coth\left(\frac{\hbar\omega}{2kT}\right) \operatorname{Im}\epsilon(r,\omega)\delta(r-r')$. Taking the expectation values at zero temperature yields $N(r,\omega) = \sqrt{4\pi\epsilon_0 \hbar \omega^2 \operatorname{Im}\epsilon(r,\omega)}$. Plugging this in, and convolving the current operator with the $\mu_0 G(r,r',\omega)$, as per the classical Maxwell equation for the vector potential, gives exactly the vector potential operator above.

**Box 2: Hamiltonians describing light–matter interactions in bound-electron systems**

We describe the Hamiltonian of light–matter interactions in bound-electron systems. A system of $N$ non-relativistic charges of masses $m_i$ and charges $q_i$, coupled to the quantized electromagnetic field is described by the *Pauli-Schrodinger* Hamiltonian $H^{PS}$:

$$H^{PS} = \sum_{i=1}^{N} \frac{\left(\boldsymbol{p}_i - q_i \boldsymbol{A}_{\text{ext}}(\boldsymbol{r}_i) - q_i \boldsymbol{A}_q(\boldsymbol{r}_i)\right)^2}{2m_i} + q_i \phi_{\text{ext}}(\boldsymbol{r}_i) + \sum_{i>j}^{N} V(\boldsymbol{r}_i, \boldsymbol{r}_j) + H_{\text{em}},$$

where $\boldsymbol{p}_i$ is the momentum operator of the $i$th particle, $\boldsymbol{r}_i$ is the corresponding position operator. $\phi_{\text{ext}}$ and $\boldsymbol{A}_{\text{ext}}$ are the scalar and vector potential of static external fields (e.g., Coulomb atomic field and a DC magnetic field). In certain cases, a strong time-dependent external field, e.g., a high intensity laser, can also be modeled as a classical field and captured by such potentials, which will become time-dependent. $\boldsymbol{A}_q$ is the quantized electromagnetic field operator. $H_{\text{em}}$ is the Hamiltonian of the electromagnetic field. $V(\boldsymbol{r}_i, \boldsymbol{r}_j)$ is the inter-particle (Coulomb) interaction, which depends on the (DC) permittivity of the medium surrounding the particles (screening), provided that electrostatic interactions with a medium are treated at a continuum, rather than atomistic level.

Commonly, the quantized photon fields have spatial variations much longer than the size of the emitter wavefunction, so that $\boldsymbol{A}_q(\boldsymbol{r}_i) \approx \boldsymbol{A}_q(0)$ (*long-wavelength approximation*), with the emitter being taken to be localized around $\boldsymbol{r} = 0$ without loss of generality. Under the long-wavelength approximation, it is possible to rigorously transform the interaction Hamiltonian to be specified in terms of the dipole moment ($\boldsymbol{d} = \sum_i \boldsymbol{d}_i$) and electric field ($\boldsymbol{E}$). This is the *dipole Hamiltonian*, given by:

$$H_{\text{dip}} = \sum_{i=1}^{N} \frac{\left(\boldsymbol{p}_i - q_i \boldsymbol{A}_{\text{ext}}(\boldsymbol{r}_i)\right)^2}{2m_i} + q_i \phi_{\text{ext}}(\boldsymbol{r}_i) + \sum_{i,j=1}^{N} V(\boldsymbol{r}_i, \boldsymbol{r}_j) + H_{\text{em}} - \boldsymbol{d} \cdot \boldsymbol{E}(0) + H_{\text{dip}}^{\text{self}},$$

where $H_{\text{dip}}^{\text{self}}$, the *dipole self-energy*, is a term whose precise form depends on how the field is quantized, but in all cases is quadratic in the dipole moment and independent of the field operators. The dipole Hamiltonian is a work-horse in atomic, molecular, and optical physics.

A key simplification arises when two levels of a bound electron system resonantly interacts with a single mode of a low-loss cavity. We may then approximate the quantized electric field in terms of a single mode, i.e., $\boldsymbol{E}(\boldsymbol{r}) \approx i\sqrt{\hbar\omega/2\epsilon_0}\,\boldsymbol{u}(0)(a - a^\dagger)$, with $\boldsymbol{u}(0)$ the cavity mode function at the emitter. We may also approximate the matter as a two-level system, i.e., $H_{\text{two-level}} = (\hbar\omega_0/2)\sigma_z$, with $\sigma_z$ the Pauli z-matrix, and $\boldsymbol{d} = \boldsymbol{d}_{fi}\sigma_x$, with $\boldsymbol{d}_{fi}$ the dipole matrix element of the two-level system. These approximations lead to the *Rabi Hamiltonian,* which is the key Hamiltonian of cavity QED:

$$H_{\text{Rabi}} = (\hbar\omega_0/2)\sigma_z + \hbar\omega a^\dagger a + \hbar\sigma_x(g^* a + g a^\dagger),$$

where we have defined $g = i\sqrt{\frac{\omega}{2\hbar\epsilon_0}}\,\boldsymbol{u}(0) \cdot \boldsymbol{d}_{fi}$. The Rabi Hamiltonian includes virtual processes in which the two-level system can be excited while also emitting a photon, as well as those in which the system can be de-excited while also absorbing a photon. If $g \ll \omega$, then these processes can be neglected under the rotating-wave approximation, reducing to the *Jaynes-Cummings Hamiltonian:*

$$H_{\text{JC}} = (\hbar\omega_0/2)\sigma_z + \hbar\omega a^\dagger a + \hbar(g\sigma^- a^\dagger + g^*\sigma^+ a),$$

with $\sigma^\pm$ being the raising (+) and lowering (-) operators of the two-level system.

**Box 3: Hamiltonians describing light–matter interactions in free-electron systems**

In general, the interaction of relativistic, spin-½ electrons with the electromagnetic field must be described by the Dirac equation. However, in many cases of interest (e.g., free electrons in microscopes and accelerators), spin weakly influences the dynamics. In such cases, the interaction can be described by the Hamiltonian of spin-less relativistic particles (*Klein-Gordon*, or *scalar QED* Hamiltonian). The corresponding Hamiltonian $H^{rel}$ given by:

$$H^{rel} = \sum_{i=1}^{N} \sqrt{m_i^2 c^4 + c^2\big(\boldsymbol{p}_i - q_i \boldsymbol{A}(\boldsymbol{r}_i)\big)^2} + q_i \phi(\boldsymbol{r}_i) + \sum_{i,j=1}^{N} V(\boldsymbol{r}_i, \boldsymbol{r}_j) + H_{\text{em}},$$
$$\approx \sum_{i=1}^{N} E(\boldsymbol{p}_i) + q_i \phi(\boldsymbol{r}_i) - \sum_{i=1}^{N} q_i \boldsymbol{A}(\boldsymbol{r}_i) \cdot \boldsymbol{v}_i + H_{\text{em}},$$

where we have approximated the square root, using the fact that the energy associated with the matter–field coupling is typically much smaller than $mc^2$. Here, we have also defined $\boldsymbol{v}_i = \boldsymbol{p}_i/m\gamma_i$, with $\gamma_i \equiv (1 - v_i^2/c^2)^{-1/2}$ being the Lorentz factor and $E(\boldsymbol{p}_i) = \sqrt{m_i^2 c^4 + c^2 \boldsymbol{p}_i^2}$ is the electron kinetic energy.

Free electrons appear to be much different than bound electrons, having a continuum of energy levels, which often necessitates including a continuum of photonic modes to describe the interaction. However, a single photonic mode can be sufficient when it is pumped by a strong driving field, or when it is *phase-matched* to the electron, i.e., $\omega_{\boldsymbol{k}} = \boldsymbol{k} \cdot \boldsymbol{v}$ for wavevector $\boldsymbol{k}$ and frequency $\omega_{\boldsymbol{k}}$. In this case, it is possible to make a single-mode approximation for the field, much like in cavity QED. Unlike conventional cavity QED, the electron cannot be approximated as a two-level system, but instead as an infinite ladder of energy levels, equally spaced in frequency by $\hbar\omega_{\boldsymbol{k}}$, and in momentum by $\boldsymbol{k}$. These are the states that are accessed by a mono-energetic electron when absorbing and emitting multiple photons in this mode. This spacing implies that the discrete energy levels of free-electron quantum emitters are *tunable* by the photon frequency. Defining a set of levels $n$ with energy $E_n = E_0 + n\hbar\omega_{\boldsymbol{k}}$, with $E_0 = \gamma mc^2$, and making a single-mode approximation, the Klein-Gordon Hamiltonian can be approximated by the *free-electron cavity QED Hamiltonian*:

$$H_{\text{CQED}}^{\text{el}} = \sum_n n\hbar\omega_{\boldsymbol{k}} |n\rangle\langle n| + \hbar\omega a^\dagger a + \hbar(g b^+ a + g^* b^- a^\dagger).$$

$b^\pm |n\rangle = |n \pm 1\rangle$, and $g = k\sqrt{1/2\hbar\epsilon_0 \omega_{\boldsymbol{k}}}\, \boldsymbol{v} \cdot \boldsymbol{u}(\boldsymbol{r_0})$, where $\boldsymbol{u}(\boldsymbol{r_0})$ is the spatial profile of the photonic quasiparticle with which the electron interacts. This simple model can explain phenomena in photon-induced near-field electron microscopy (PINEM) [Kfir2019, DiGiulio2019].

[Endnote1] That the spontaneous emission is proportional to the imaginary part of the Green's function is a manifestation of the fact that spontaneous emission can be seen as emission driven or "stimulated" by vacuum fluctuations of the quantized electromagnetic field. In particular, the fluctuations of the quantized electric field, given by $\langle 0|E_i(\boldsymbol{r},\omega)E_j(\boldsymbol{r}',\omega)|0\rangle$, with $|0\rangle$ the vacuum state of the field, are related to the Green's function via the fluctuation-dissipation relation through $\langle 0|E_i(\boldsymbol{r},\omega)E_j(\boldsymbol{r}',\omega')|0\rangle = \frac{\mu_0}{\pi}\hbar\omega^2 \, \text{Im} \, G_{ij}(\boldsymbol{r},\boldsymbol{r}',\omega)\delta(\omega-\omega')$. More complex phenomena than single-photon spontaneous emission, such as multi-photon spontaneous emission and vacuum energy shifts, are also related to vacuum fluctuations. Consequently, dependences on the imaginary part of the Green's function are ubiquitous in light–matter interactions.

[Endnote2] The Rabi splitting in these macroscopic experiments in some sense are quite classical, as the Rabi splitting can be *quantitatively* calculated in these many-molecule experiments purely through classical physics. In particular, the Rabi splitting can be obtained by modeling the molecular assembly as a Lorentz oscillator of the appropriate geometry, and solving Maxwell's equations for resonance frequencies in the system of this Lorentz oscillator coupled to the dielectric or metal materials constituting the cavity.

[Endnote3] The photonic quasiparticle vacuum changes the energy levels of emitters in a way that in principle depend on *all* modes, even arbitrarily high frequency ones; inviting questions as to how to find correct predictions for energy shifts. See [Tsytovich1955] for the case of a homogeneous media.